\documentclass{iopart}
\usepackage{graphicx}
\input{epsf}
\usepackage{amssymb}
\begin{document}

\title{Towards two-dimensional search engines} 
\author{L. Ermann$^1$, A.D. Chepelianskii$^2$ and D.L. Shepelyansky$^1$}
\address{$^1$Laboratoire de Physique Th\'eorique du CNRS, IRSAMC, 
Universit\'e de Toulouse, UPS, 31062 Toulouse, France}
\address{$^2$ Cavendish Laboratory, Department of Physics, 
University of Cambridge, CB3 0HE, United Kingdom}
% \eads{\mailto{$^1$ermann@irsamc.ups-tlse.fr}, \mailto{$^2$dima@irsamc.ups-tlse.fr}}

%\date{\today}
%\date{Received  30 June 2011, 29 February 2012}

\begin{abstract}
We study the statistical properties of various directed networks
using ranking of their nodes based on
the dominant vectors of the Google matrix
known as PageRank and CheiRank.
On average PageRank orders nodes 
proportionally to a number of ingoing links,
while CheiRank orders nodes proportionally to
a number of outgoing links. In this way
the ranking of nodes becomes two-dimensional
that paves the way for development of
two-dimensional search engines of new type.
Statistical properties of information flow on PageRank-CheiRank plane
are analyzed for networks of British, French and Italian Universities,
Wikipedia,  Linux Kernel, gene regulation and other networks.
A special emphasis is done for British Universities networks
using the large  database publicly available at UK.
Methods of spam links control are also analyzed.
\end{abstract}

% {World Wide Web, Internet}
% }

\submitto{\JPA}
\maketitle

\section{Introduction}
 
 During the last decade the modern society developed enormously
 large communication networks. The well known example
 is the World Wide Web (WWW) which starts to approach to
 $10^{11}$ webpage \cite{websize}. The sizes of social networks 
 like Facebook \cite{facebook} and VKONTAKTE  \cite{vkontakte}
 also become enormously large reaching 600 and 100 millions user pages
 respectively. The information retrieval from such huge data
 bases becomes the foundation and main  challenge for search engines
 \cite{searchengwiki,searcheng}. The fundamental basis of Google search engine
 is the PageRank algorithm \cite{brin}. This algorithm ranks all websites 
 in a decreasing order of  components of the PageRank vector
 (see e.g. detailed description at \cite{meyerbook}, historical surveys
 of PageRank are given at \cite{franceschet,vigna}). 
 This vector is a right eigenvector of the Google matrix at the unit eigenvalue, it is
 constructed on the basis of the adjacency matrix of the directed network,
 its components give a probability to find a random surfer on a given node.
\begin{figure} 
\begin{indented}\item[]
\begin{center}
\includegraphics[width=0.36\textwidth]{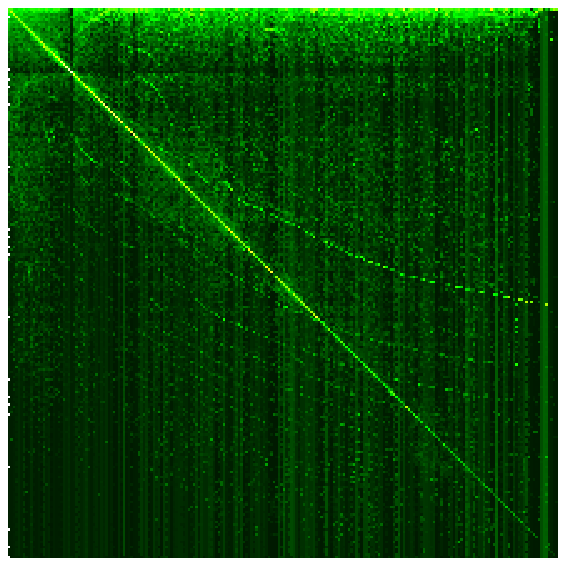}
\includegraphics[width=0.36\textwidth]{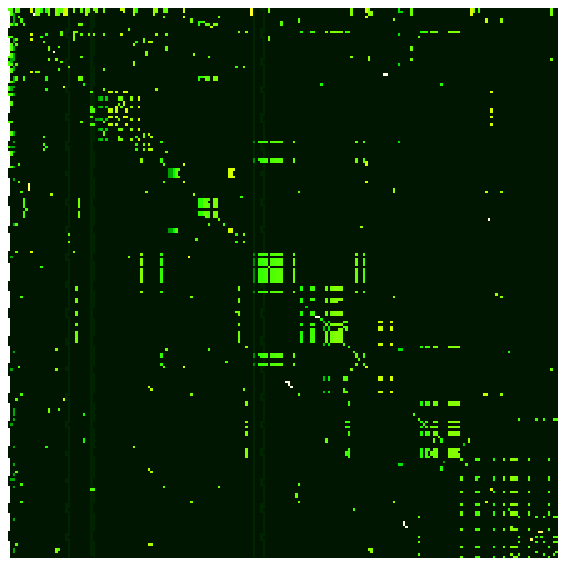}\\
\includegraphics[width=0.36\textwidth]{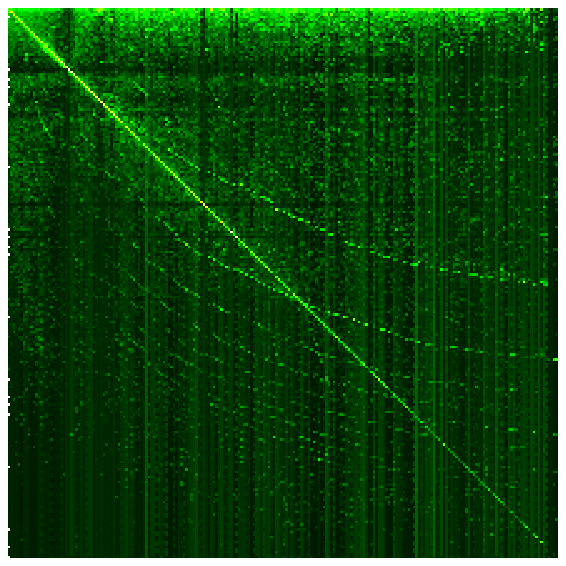}
\includegraphics[width=0.36\textwidth]{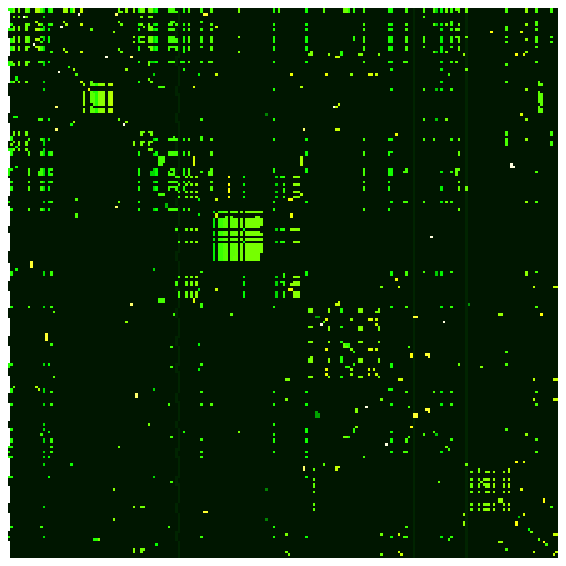}\\
\includegraphics[width=0.36\textwidth]{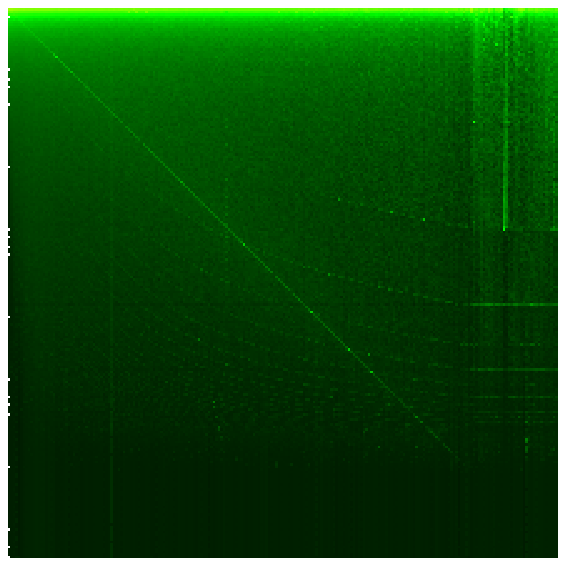}
\includegraphics[width=0.36\textwidth]{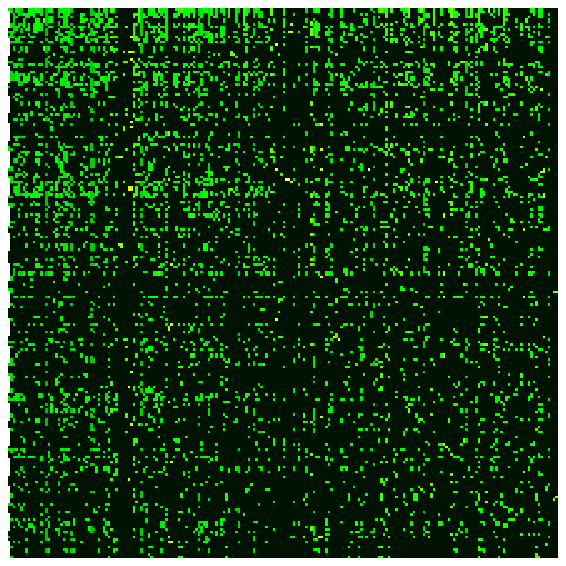}\\
\includegraphics[width=0.36\textwidth]{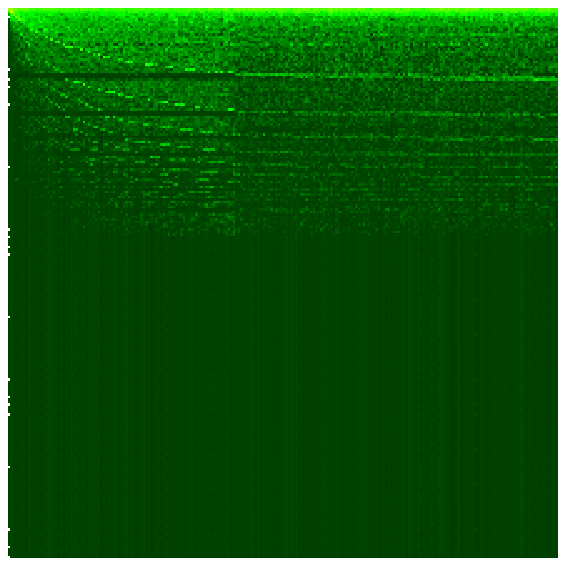}
\includegraphics[width=0.36\textwidth]{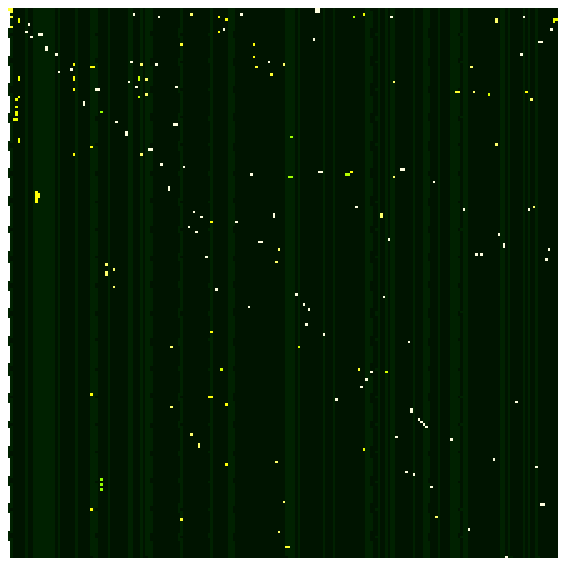}
\end{center}
\vglue -0.4cm
\caption{Left column: coarse-grained density of Google matrix elements
$G_{i,j}$  
written in the PageRank basis $K(i)$ 
with indexes $j \rightarrow K(i)$ (in $x$-axis) and $i \rightarrow K'(i)$ 
(in a usual matrix representation with $K=K'=1$ on top left corner);
the coarse graining is done on $500 \times 500$ square cells 
for the networks of University of Cambridge 2006,
University of Oxford 2006,
Wikipedia English articles,
PCN of Linux Kernel V2.6 (from top to bottom).
Right column shows the first  $200 \times 200$
matrix elements of $G$ matrix at  $\alpha=0.85$
without coarse graining with the
same order of panels as in the left column.
Color shows the density of matrix elements
changing from black for minimum value ($(1-\alpha)/N$ to white for maximum
value via green and yellow 
(density is coarse-grained in left column).
 All  matrices are shown in the basis of
PageRank index $K$ (and $K'$) of matrix $G_{KK'}$, 
which corresponds to $x$ (and $y$) 
axis with $1 \leq K,K' \leq N$ (left column)
and $1 \leq K,K' \leq 200$ (right column); 
all nodes are ordered
by PageRank index $K$ of matrix $G$ and thus we have two matrix indexes
$K,K'$ for matrix elements in this basis.
}
\label{fig1}
\end{indented}
\end{figure} 

The Google matrix $G$ of 
a directed network with $N$ nodes is given by
\begin{equation}
  G_{ij} = \alpha S_{ij} + (1-\alpha) / N \;\; ,
\label{eq1} 
\end{equation} 
where the matrix $S$ is obtained by normalizing 
to unity all columns of the adjacency matrix $A_{i,j}$,
and replacing columns with zero elements by $1/N$.
An element $A_{ij}$ of the adjacency matrix is equal to unity
if a node $j$ points to node $i$ and zero otherwise.
The damping parameter $\alpha$ in the WWW context describes the probability 
$(1-\alpha)$ to jump to any node for a random surfer. 
The value $\alpha = 0.85$  gives
a good classification for WWW \cite{meyerbook}
and  thus we also use this value here.
A few examples of Google matrix for various directed networks
are shown in Fig.~\ref{fig1}.
The matrix $G$ belongs to the class of Perron-Frobenius 
operators \cite{meyerbook},
its largest eigenvalue 
is $\lambda = 1$ and other eigenvalues have $|\lambda| \le \alpha$. 
The right eigenvector 
at $\lambda = 1$ gives the probability $P(i)$ to find 
a random surfer at site $i$ and
is called the PageRank. Once the PageRank is found, 
all nodes can be sorted by decreasing probabilities $P(i)$. 
The node rank is then given by index $K(i)$ which
reflects the  relevance of the node $i$. 
The PageRank dependence on $K$ is well
described by a power law $P(K) \propto 1/K^{\beta_{in}}$ with
$\beta_{in} \approx 0.9$. This is consistent with the relation
$\beta_{in}=1/(\mu_{in}-1)$ corresponding to the average
proportionality of PageRank probability $P(i)$
to its in-degree distribution $w_{in}(k) \propto 1/k^{\mu_{in}}$
where $k(i)$ is a number of ingoing links for a node $i$  
\cite{litvak,meyerbook}. For the WWW it is established that
for the ingoing links $\mu_{in} \approx 2.1$ (with $\beta_{in} \approx 0.9$)
while for out-degree distribution
$w_{out}$ of
outgoing links a power law has the exponent  $\mu_{out} \approx 2.7$
\cite{donato,upfal}.
Similar values of these exponents are found for 
the WWW British university networks \cite{ggsuniv}, 
the procedure call network 
(PCN) of Linux Kernel software introduced in \cite{alik}
and for Wikipedia hyperlink citation network of English articles 
(see e.g. \cite{wiki}). 

The PageRank gives at the top the most known and popular nodes.
However, an example of the Linux PCN studied in \cite{alik}  
shows that in this case the PageRank puts at the top certain procedures
which are not very important from the software view point
(e.g. {\it printk}).
As a result it was proposed \cite{alik} to use
in addition another ranking taking the network
with inverse link directions in the adjacency matrix
corresponding to $A_{ij} \rightarrow A^T=A_{ji}$
and constructing from it an additional Google matrix $G^*$
according to relation (\ref{eq1}) at the same $\alpha$. 
The eigenvector of $G^*$ with eigenvalue 
$\lambda=1$ gives then a new inverse PageRank $P^*(i)$
with ranking index $K^*(i)$. This ranking was named 
CheiRank \cite{wiki} to mark that it allows to
{\it chercher l'information} in a new way
(that in English means {\it search the information} in a new way).
Indeed, for the Linux PCN the CheiRank gives at the top
more interesting and important procedures compared to 
the PageRank \cite{alik} (e.g. {\it start\_kernel}). 
While the PageRank ranks the network nodes in average 
proportionally to a number of ingoing links, the CheiRank 
ranks nodes in average proportionally to a number of outgoing links. 
The physical meaning of PageRank vector components is 
that they give the probability 
to find  a random surfer on a given node when 
a surfer follows the given directions of network links.
In a similar way the 
CheiRank vector components give the probability 
to find  a random surfer on a given node when 
a surfer follows the inverted directions of network links.
Since each node belongs both to CheiRank and PageRank vectors 
the ranking of information flow on a directed network 
becomes {\bf two-dimensional}. We note that
there have been earlier studies of
PageRank of the Google matrix with inverted directions of links
\cite{fogaras,hristidis}, but no systematic analysis of 
statistical properties of 2DRanking was presented there.

\begin{figure} 
\begin{indented}\item[]
\begin{center}
\includegraphics[width=0.67\textwidth]{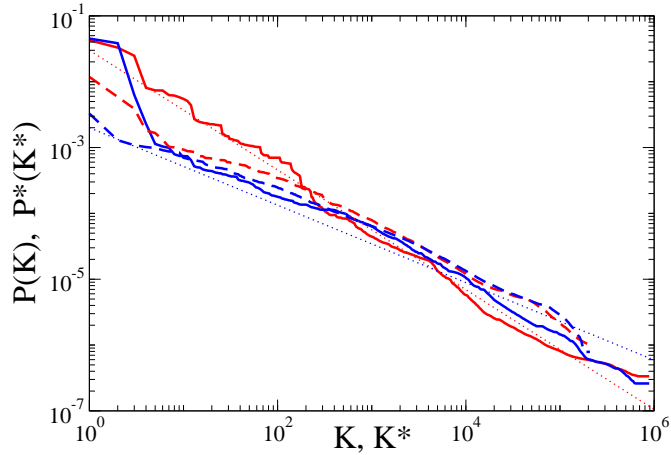}
\end{center}
\vglue -0.0cm
\caption{Dependence of probabilities of
PageRank $P(K)$ (red curve) and CheiRank $P^*(K^*)$ (blue curve)
on corresponding ranks $K$ and $K^*$
for the network of University of Cambridge in 2006
(dashed curve) and in 2011 (full curve).
The power law dependencies with the
exponents $\beta \approx 0.91; 0.59$,
corresponding to the relation 
$\beta=1/(\mu-1)$ with $\mu=2.1; 2.7$
respectively, are shown by dotted straight lines.
}
\label{fig2}
\end{indented}
\end{figure}

An example of variation of PageRank probability $P(K)$ with $K$
and CheiRank probability   $P^*(K^*)$ with $K^*$
are shown in Fig.~\ref{fig2} for the WWW network of University of Cambridge in
years 2006 and 2011. Other examples for PCN Linux Kernel and
Wikipedia can be find in \cite{alik,wiki}. Detailed parameters of networks
which we analyze in this paper and their sources are given in Appendix. 

A detailed comparative analysis of PageRank 
and $\;\;$ CheiRank two-dimensional classification
was done in \cite{wiki} on the example
of Wikipedia hyperlink citation network of English articles. 
It was shown that CheiRank highlights communicative property
 of nodes leading to a new way of two-dimensional ranking. 
While according to the PageRank top three countries are 
{\it 1. USA, 2. UK, 3. France} the CheiRank gives 
{\it 1.India, 2.Singapore, 3.Pakistan} 
as most communicative Wikipedia country articles. 
Top 100 personalities of PageRank has  the following percents  
in 5 main category activities 
58 (politics), 10 (religion), 17 (arts), 15 (science), 0 (sport)
 \cite{wiki}. 
Clearly the significance of politicians is overestimated. 
In contrast, the CheiRank gives more balanced distribution 
over these categories with 15, 1, 52, 16, 16 
respectively. It allows to classify information 
in a new way  finding composers, architects, botanists, 
astronomers who are not well known but who, for example,  
discovered  a lot of Australians butterflies ({\it George Lyell}) 
or many asteroids ({\it Nikolai Chernykh}). 
These two persons appear in the large Listings of  Australians butterflies
and in the  Listing of Asteroids (since they discovered many of them)
and due to that they gain high CheiRank values. 
This shows that the information retrieval, which uses 
both PageRank and CheiRank, allows to rank nodes
not only by an amount of their popularity
(how known is a given node)
but also by an amount of their communicative property
(how communicative is a given node).
This 2DRanking 
was also applied to the brain model of neuronal network \cite{brain}
and the business process management network \cite{business}
and it was shown that it gives a new useful way of 
information treatment in these networks. 
The 2DRanking in the PageRank-CheiRank plane 
also naturally appears 
for the world trade network corresponding to 
import and export trade flows \cite{wtrade}.
Thus the 2DRanking
based on PageRank and CheiRank paves the way 
to a development of 2D search engines 
which can become more intelligent 
than the present Google search based on 1D PageRank algorithm. 

In this work we study the statistical properties of such a 2DRanking
using examples of various real directed networks including the WWW
of British, French and Italian University networks 
\cite{britishuniv}, Wikipedia network \cite{wiki}, 
Linux Kernel networks \cite{alik,wlinux}, 
gene regulation networks \cite{gene,ualon}
and other networks. The paper is constructed as following:
in Section 2 we study the properties of node density
in the plane of PageRank and CheiRank,
in Section 3 the correlator properties between PageRank and 
CheiRank vectors are analyzed for various networks,
information flow on the plane of PageRank and CheiRank is 
analyzed in Section 4, 
methods of control of SPAM outgoing links are discussed in 
Section 5,
2DRanking applications for the gene regulation networks
are considered in Section 6, discussion of results is 
presented in Section 7. The parameters of the networks 
and references on their sources are given in Appendix.

\section{Node Density of 2DRanking}

\begin{figure} 
\begin{indented}\item[]
\begin{center}
\includegraphics[width=0.37\textwidth]{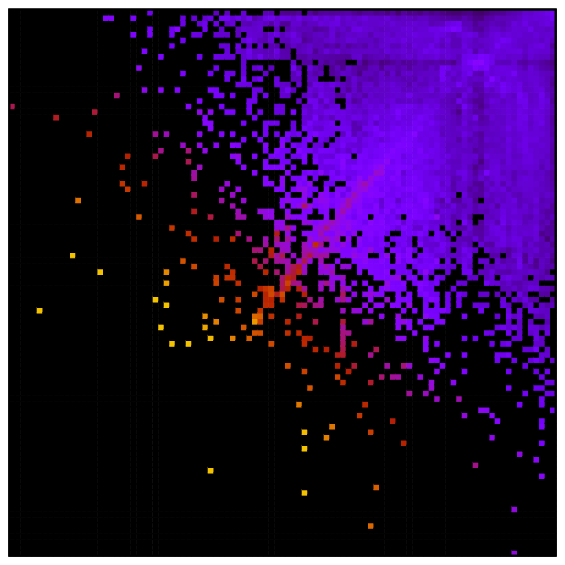}
\includegraphics[width=0.37\textwidth]{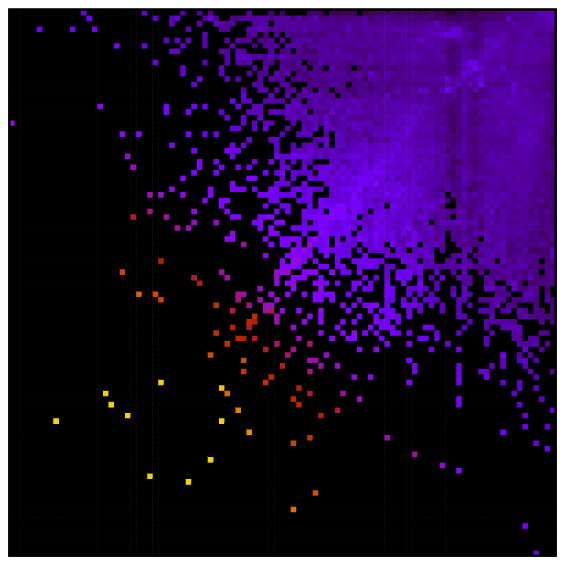}\\
\includegraphics[width=0.37\textwidth]{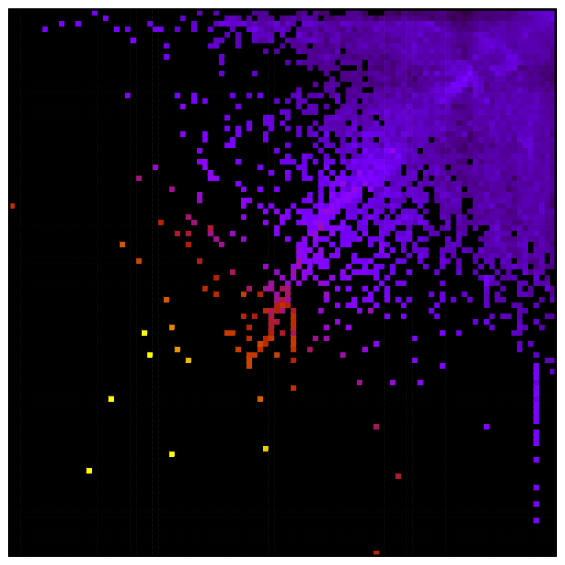}
\includegraphics[width=0.37\textwidth]{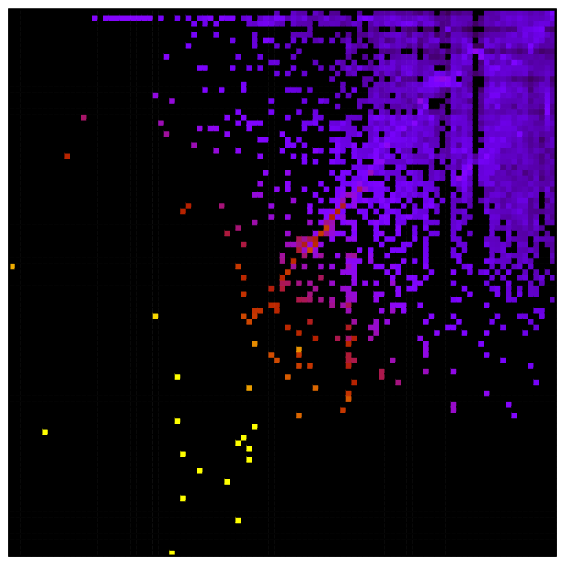}\\
\includegraphics[width=0.42\textwidth]{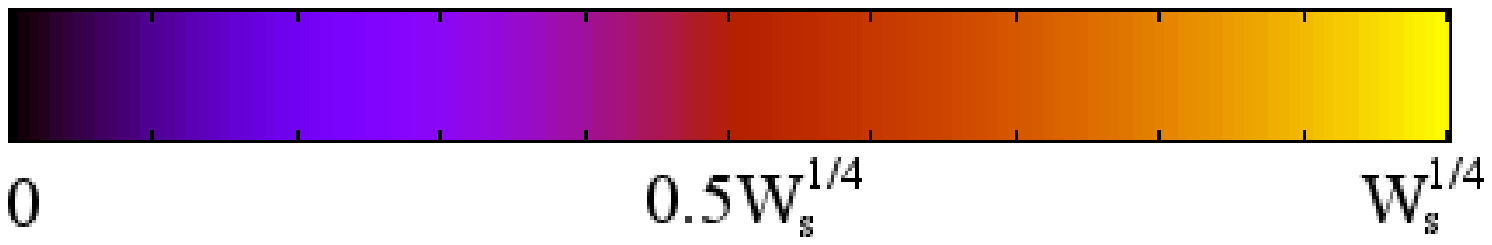}
\end{center}
\vglue -0.0cm
\caption{Density distribution $W(K,K^*)=dN_i/dKdK^*$ 
for  networks of four British Universities
in the plane of PageRank $K$ and CheiRank $K^*$ 
indexes in log-scale $(\log_N K,\log_N K^*)$.
The density is shown for 
$100\times100$ equidistant grid in $\log_N K,\log_N K^*\in[0,1]$,
the density is averaged over all nodes inside each cell of the grid,
the normalization condition is $\sum_{K,K^*}W(K,K^*)=1$.
Color varies from black for zero to yellow 
for maximum density value $W_M$ with a saturation value of 
$W_s^{1/4}=0.5W_M^{1/4}$ so that the same color
is fixed for $0.5W_M^{1/4} \leq  W^{1/4} \leq W_M^{1/4}$
to show in a better way low densities.
The panels  show networks of University of Cambridge (2006)
with $N=212710$ (top left);
University of Oxford with $N=200823$ (top right);
University of Bath with $N=73491$ (bottom left);
University of East Anglia with $N=33623$ (bottom right).
The axes show: $\log_N K$ in $x$-axis,  $\log_N K^*$ in $y$-axis,
in both axes the variation range is $(0,1)$. 
}
\label{fig3}
\end{indented}
\end{figure} 

A few examples of the Google matrix for four directed networks 
are shown in Fig.~\ref{fig1}. There is a significant similarity in
the global structure of $G$ for Universities of Cambridge and Oxford
with well visible hyperbolic curves (left column) even if at small scales 
the matrix elements are rather different 
(right column) in these two networks
(see Fig.~\ref{fig1}). Such hyperbolic curves are also visible
in the Google matrix of Wikipedia (left column)
even if here they are less pronounced due to 
much larger averaging inside the cells which
contain about $15$ 
times larger number of nodes (see network parameters in Appendix).
We make a conjecture that the appearance of such curves is related to
existence of certain natural categories existing in the network,
e.g. departments for Universities or countries, cities, personalities
etc for Wikipedia. We expect that there are relatively more links inside
a given category compared to links between categories.
However, this is only a statistical property since on small
scales at small $K$ values 
the hyperbolic curves are not visible  (right column in Fig.~\ref{fig1}).
Hence, more detailed studies are required for
verification of this conjecture.
At small scale $G$ matrix of Wikipedia is much more dense
compared to the cases of Cambridge and Oxford (right column).
We attribute such an increase of density of 
significant matrix elements to a stronger connectivity
between nodes with large $K$ in Wikipedia compared to the 
case of universities where the links have more hierarchical structure.
Partially this increase of density can be attributed to 
a larger number of links per node in the case of Wikipedia
but this increase by a factor $2.1$ is not so strong and cannot
explain all the differences of densities at small $K$ scale.
For Wikipedia there is about 20\% of nodes 
at the bottom of the matrix where there are almost no links.
For PCN of Linux Kernel this fraction becomes significantly
larger with about 60\% of nodes.
The hyperbolic curves are still well visible for Linux PCN
inside remaining 40\% of nodes. On a small scale the density
of matrix elements for Linux is rather small
compared to the three previous cases. 
We attribute this to a much smaller number of links per node
which is by factor 5 smaller for Linux compared to the university
networks of Fig.~\ref{fig1} (see data in Appendix).  

\begin{figure} 
\begin{indented}\item[]
\begin{center}
\includegraphics[width=0.37\textwidth]{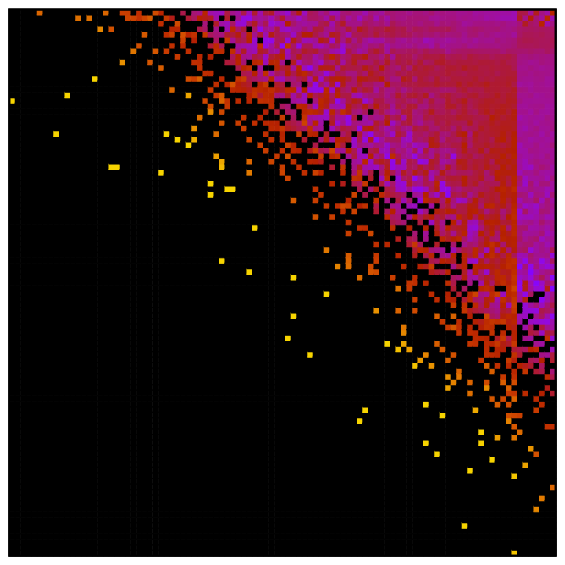}
\includegraphics[width=0.37\textwidth]{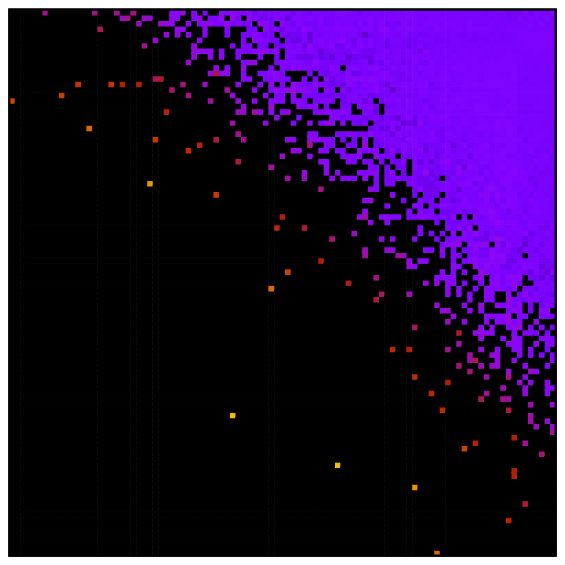}\\
\includegraphics[width=0.37\textwidth]{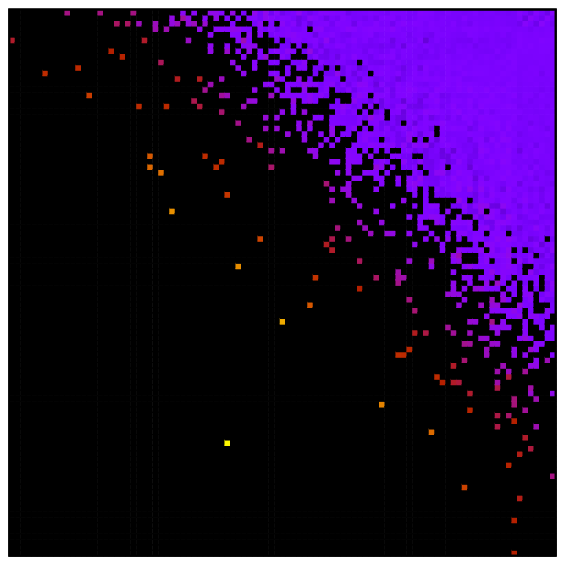}
\includegraphics[width=0.37\textwidth]{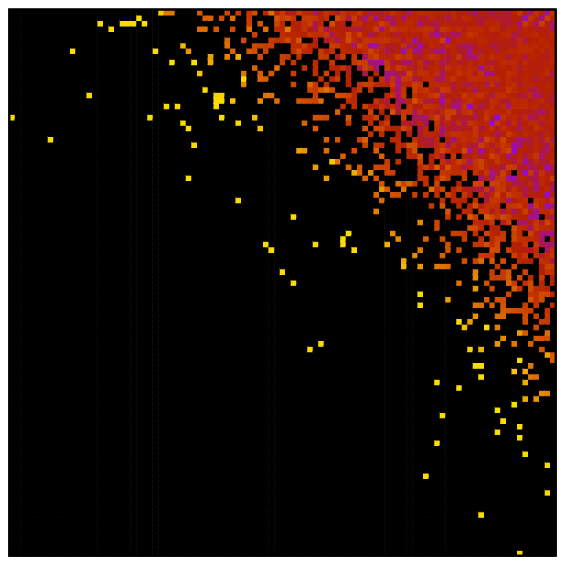}\
\end{center}
\vglue 0.0cm
\caption{Density distribution $W(K,K^*)=dN_i/dKdK^*$ 
of four Linux Kernel networks
shown in the same frame as in Fig.~\ref{fig3}.
The panels show networks for Linux versions
V2.0 with $N=14080$ (top left);
V2.3 with $N=41117$ (top right);
V2.4 with $N=85757$ (bottom left);
V2.6 with $N=285510$ (bottom right).
Color panel is the same as in Fig.~\ref{fig3}
with a saturation value of 
$W_s^{1/4}=0.2W_M^{1/4}$ so that the same color
is fixed for $0.2W_M^{1/4} \leq  W^{1/4} \leq W_M^{1/4}$
to show in a better way low densities.
The axes show: $\log_N K$ in $x$-axis,  $\log_N K^*$ in $y$-axis,
in both axes the variation range is $(0,1)$.
}
\label{fig4}
\end{indented}
\end{figure} 

The distributions of density of nodes $W(K,K^*)=dN_i/dKdK^*$
in the plane of PageRank and CheiRank
in log-scale are shown for four networks of British
Universities in Fig.~\ref{fig3}. 
Here $d N_i$ is a number of nodes in a cell of size
$dKdK^*$ (see detailed description in \cite{wiki}).
Even if the coarse-grained
$G$ matrices for Cambridge and Oxford look rather similar
the density distributions in $(K,K^*)$ plane
are rather different.
The density distributions  for all four universities
clearly show 
that nodes with high PageRank have low CheiRank
that corresponds to zero density at low $K$, $K^*$
values. At large $K$, $K^*$ values there is a maximum line
of density which is located not very far from 
the diagonal $K \approx K^*$. The presence of such 
a line should correspond to significant correlations
between $P(K(i))$ and  $P^*(K^*(i))$ vectors
that will be discussed in more detail in next Section. 
The presence of correlations between $P(K(i))$ and  $P^*(K^*(i))$
leads to a probability distribution with one main
maximum along a diagonal at $K+K^*=const$.
This is similar to the properties of density distribution
for the Wikipedia network discussed in \cite{wiki}
(see also bottom right panel in Fig.~\ref{fig13} below).

The density of nodes for  Linux  networks is shown
in Fig.~\ref{fig4}. In these networks the density is
homogeneous along lines $K+K^*=const$
that corresponds to absence of correlations
between  $P(K(i))$ and  $P^*(K^*(i))$.
Indeed, in absence of such correlations
the distribution of nodes in $K$, $K^*$ plane
is given by the product of independent probabilities.
In the log-scale format used in Fig.~\ref{fig4} 
this leads to a homogeneous density of nodes in the top right
corner of $(\log_N K, \log_N K^*)$ plane as it was discussed in
\cite{wiki} (see right panel in Fig.~4 there).
Indeed, the distributions in Fig.~\ref{fig4}
are very homogeneous inside  top-right triangle.
We note that, a part of fluctuations,
the distributions remain rather stable even if the size
of the network is changed by factor 20
from V2.0 to V2.6 version. 
The physical reasons for absence of correlations 
between $P(i)$ and $P^*(i)$ have been explained in \cite{alik}  
on the basis of the concept of "separation of concerns" 
used in software architecture. As discussed in  \cite{alik},
a good code should decrease a number of  procedures that have
high values of both PageRank and CheiRank  
since such  procedures will play a
critical role in error propagation since they are both
popular and highly communicative at the
same time. For example in the Linux Kernel, {\it do\_fork()},
that creates new processes, belongs to this class. These
critical procedures may introduce subtle errors because
they entangle otherwise independent segments of code.
The above observations suggest that the independence
between popular procedures, which have
high $P(K_i)$ and fulfill important but well
defined tasks, and  communicative procedures, which 
have high $P^*(K^*_i)$ and organize
and assign tasks in the code, is an important ingredient 
of well structured software. 
  We discuss the properties 
of PageRank-CheiRank correlations
in the next Section.

\section{Correlations between PageRank and CheiRank}

The correlations between PageRank and CheiRank can be quantitatively 
characterized by the correlator
\begin{equation}
  \kappa(\tau) =N \sum^N_{i=1} P(K(i)+\tau) P^*(K^*(i)) - 1 \;\; .
\label{eq2} 
\end{equation}
Such a correlator was introduced in \cite{alik}
for $\tau=0$ and we will use the same notation
$\kappa=\kappa(\tau=0)$. This correlator at $\tau=0$
shows if there are correlations and dependencies
between PageRank and CheiRank vectors.
Indeed, for homogeneous vectors $P(K)=P^*(K^*)=1/N$
we have $\kappa=0$ corresponding to absence of correlations.
We will see below that the values of $\kappa$
are very different for various directed networks.
Hence, this new characteristic is able to distinguish
various types of networks even if  they have rather similar
algebraic decay of PageRank and CheiRank vectors.   
\begin{figure} 
\begin{indented}\item[]
\begin{center}
\includegraphics[width=0.67\textwidth]{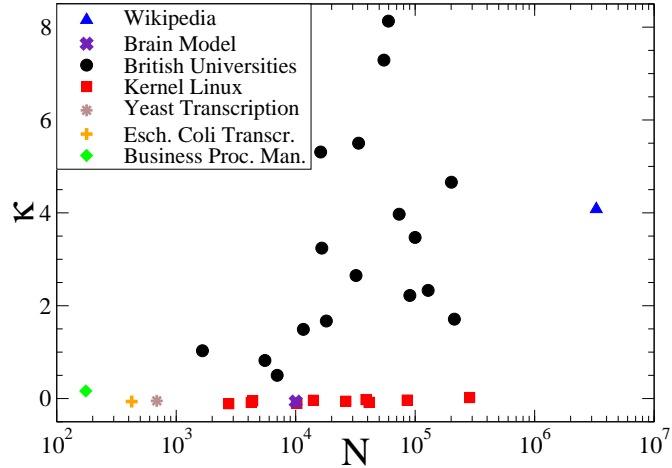}
\end{center}
\vglue 0.0cm
\caption{Correlator $\kappa$ as a function of the number of nodes $N$
for different networks:  Wikipedia network, 
17 British Universities, 10 versions of Kernel 
Linux Kernel PCN, 
Escherichia Coli and Yeast Transcription Gene networks, 
Brain Model Network and
Business Process Management Network.
The parameters of networks are given in Appendix.
}
\label{fig5}
\end{indented}
\end{figure} 

The values of $\kappa$
for networks of various size $N$ are shown in Fig.~\ref{fig5}.
The two types of networks are well visible
according to these data. The human created 
university and Wikipedia networks have typical values of
$\kappa$ in the range $1 < \kappa < 8$.
Other networks like Linux PCN, 
Gene Transcription networks, brain model and business process 
management networks have $\kappa \approx 0$.

\begin{figure} 
\begin{indented}\item[]
\begin{center}
\includegraphics[width=0.67\textwidth]{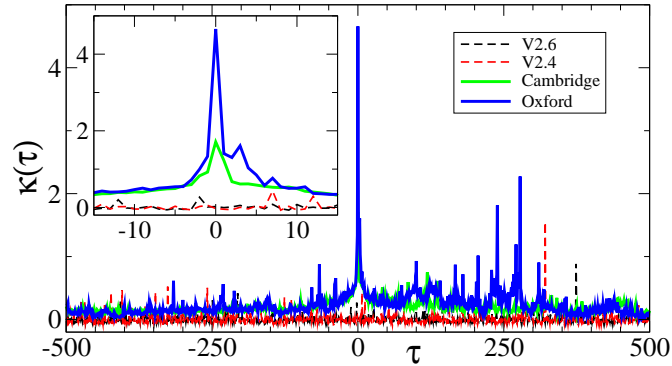}
\end{center}
\vglue 0.0cm
\caption{Correlator $\kappa(\tau)$ for two different long and short range of 
$\tau$ in the main and inset panel respectively.
The Kernel Linux PCN $V2.6$ and $V2.4$ are shown by dashed curves while  
and Universities networks of Cambridge and Oxford are shown by full curves.
}
\label{fig6}
\end{indented}
\end{figure} 

The dependence of $\kappa(\tau)$ on the correlation ``time''
$\tau$ is shown in Fig.~\ref{fig6}. For the PCN of Linux there are no 
correlations at any $\tau$ while for the university
networks we find that the correlator drops to small values
with increase of $|\tau|$ (e.g. $|\tau| > 5$)
even if at certain rather large values of  $|\tau|$
significant values of correlator $\kappa$ can reappear.

\begin{figure} 
\begin{indented}\item[]
\begin{center}
\includegraphics[width=0.67\textwidth]{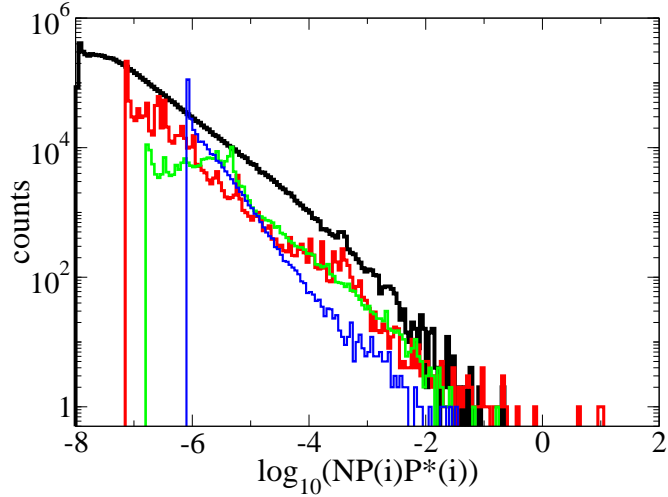}
\end{center}
\vglue 0.0cm
\caption{Histogram of frequency appearance of   
correlator components $\kappa_i=N P(K(i)) P^*(K^*(i))$ 
for networks of Wikipedia (black), University of
Cambridge in 2006 (green) and in 2011 (red),  
and PCN of Linux Kernel V2.6 (blue).
For the histogram 
the whole interval  $10^{-8} \leq \kappa_i \leq 10^2$ is divided in 
200 cells of equal size in logarithmic scale.
}
\label{fig7}
\end{indented}
\end{figure} 

It is interesting to see what are typical values
$\kappa_i=N P(K(i)) P^*(K^*(i))$ of contributions
in the correlator sum (\ref{eq2}) at $\tau=0$.
The distribution of $\kappa_i$ values for 
a few networks are shown in Fig.~\ref{fig7}.
All of them follow a power law
with an exponent $a=$1.23 for PCN Linux,
$0.70 $ Wikipedia and Univ. of Cambridge $0.76$ (2006)
and $0.66$ (2011).
We note that further studies are required to
obtain analytically the values of the exponent $a$.
In the later two cases the exponent and the distribution shape
remains stable in time, however,
in 2011 there appear few nodes
with very large $\kappa_i$ values
which give a significant increase of the 
correlator from $\kappa=1.71$ (in 2006) 
up to $\kappa=30.0$ (in 2011).  It is possible that
such a situation can appear if it is imposed that
practically any page points to the main university page
which may have rather high CheiRank 
due to many outgoing links to other
departments and university divisions.
We suppose that these are also the reasons due to which
we have appearance of large values of 
 $\kappa(\tau)$ in University networks.
At the same time more detailed studies are required to
clarify the correlation properties 
on directed networks of a deeper level.
We will return to a discussion of
university networks collected in 2011 in Section 7.

\begin{figure} 
\begin{indented}\item[]
\begin{center}
\includegraphics[width=0.67\textwidth]{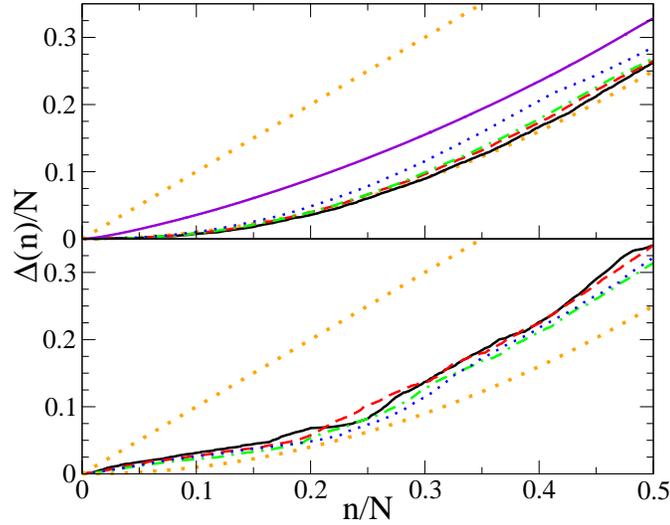}
\end{center}
\vglue 0.0cm
\caption{Dependence of the point-count correlation function 
$\Delta(n)/N$ on $n/N$
for networks of 
Wikipedia, British Universities, and Kernel Linux PCN.
The curves in the top panel show the cases of 
Wikipedia (solid violet) and four versions of
PCN of Linux Kernel with $V2.0$ (solid black), 
$V2.3$ (dashed red), 
$V2.4$ (dot-dashed green), 
and $V2.6$ (dotted blue).
The curves in the bottom panel show 
the cases of British Universities with 
East Anglia (solid black), Bath (dashed red), 
Oxford (dot-dashed green), and Cambridge 2006 
(dotted blue).
Dotted orange curves represent the totally correlated case 
with $\Delta(n)/N=n/N$, and the totally uncorrelated one with
$\Delta(n)/N=(n/N)^2$.
}
\label{fig8}
\end{indented}
\end{figure} 

Another way to analyze the correlations between
PageRank and CheiRank is simply to 
count the number of nodes $\Delta(n)$
inside a square $1 \leq K(i), K^*(i) \leq n$.
For a totally correlated distribution
with $K(i)=K^*(i)$
we  have $\Delta(n)/N=n/N$ while in
absence of correlations we should have points
homogeneously distributed inside a square $n \times n$
that gives $\Delta(n)/N=(n/N)^2$. 
The dependence of such point-count correlator
$\Delta(n)$ on size $n$ is displayed in Fig.~\ref{fig8}
for various networks. These data clearly show that the Linux PCN 
is uncorrelated being close to the limiting
uncorrelated dependence while Wikipedia
and British University networks show intermediate strength 
of correlations being between the two limiting
functions of $\Delta(n)$.

\section{Information flow of 2DRanking}

\begin{figure} 
\begin{indented}\item[]
\begin{center}
\includegraphics[width=0.37\textwidth]{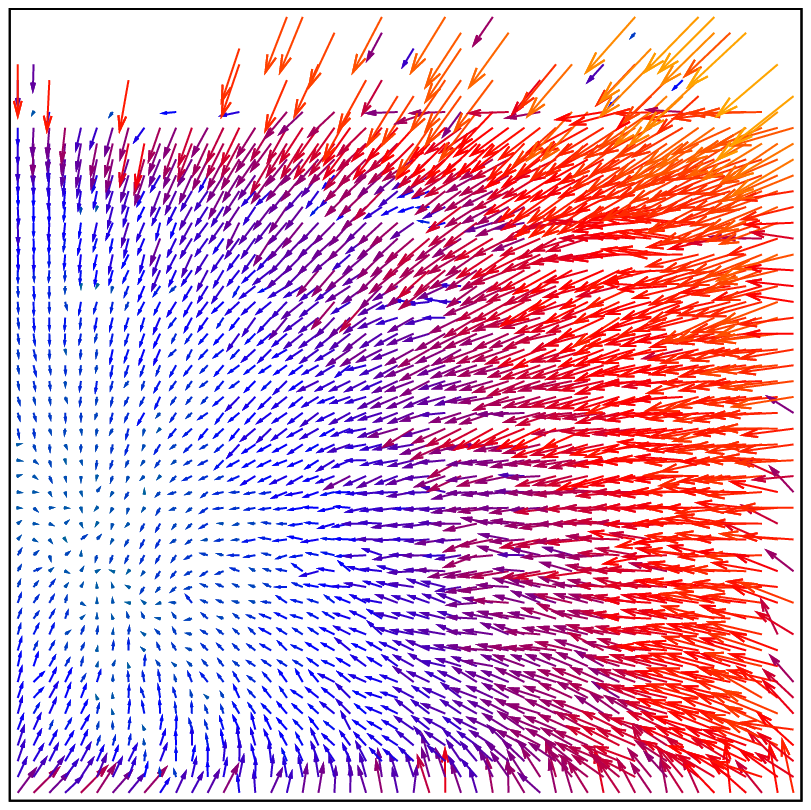}
\includegraphics[width=0.37\textwidth]{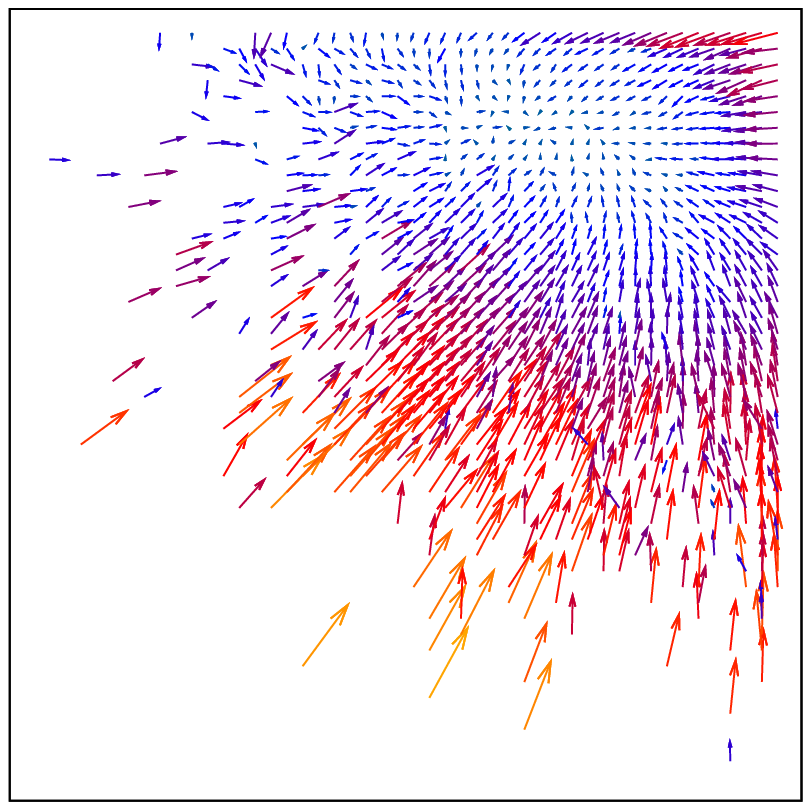}\\
\includegraphics[width=0.37\textwidth]{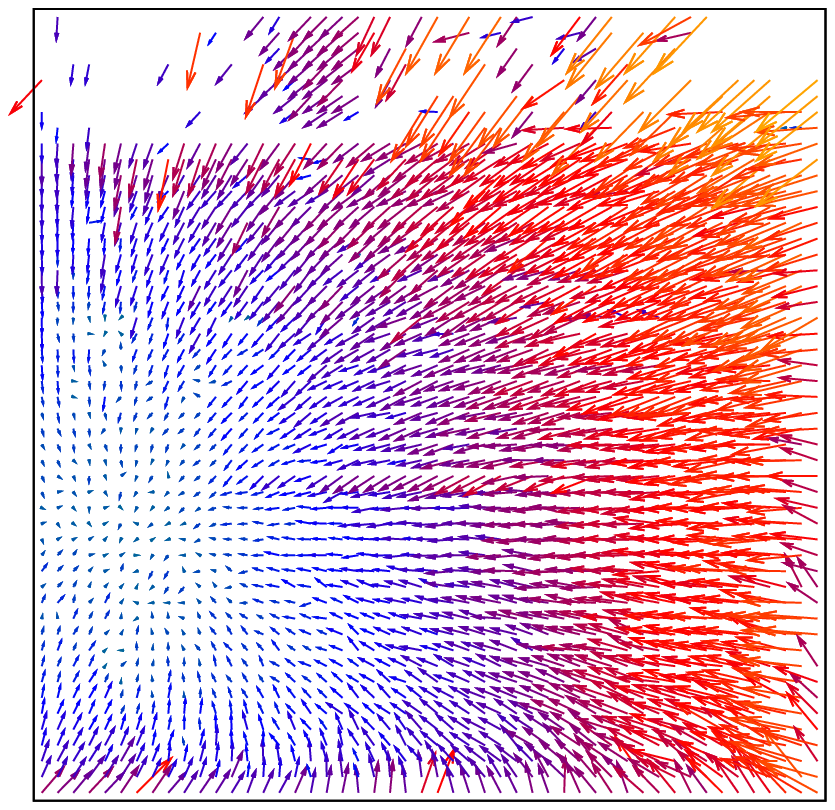}
\includegraphics[width=0.37\textwidth]{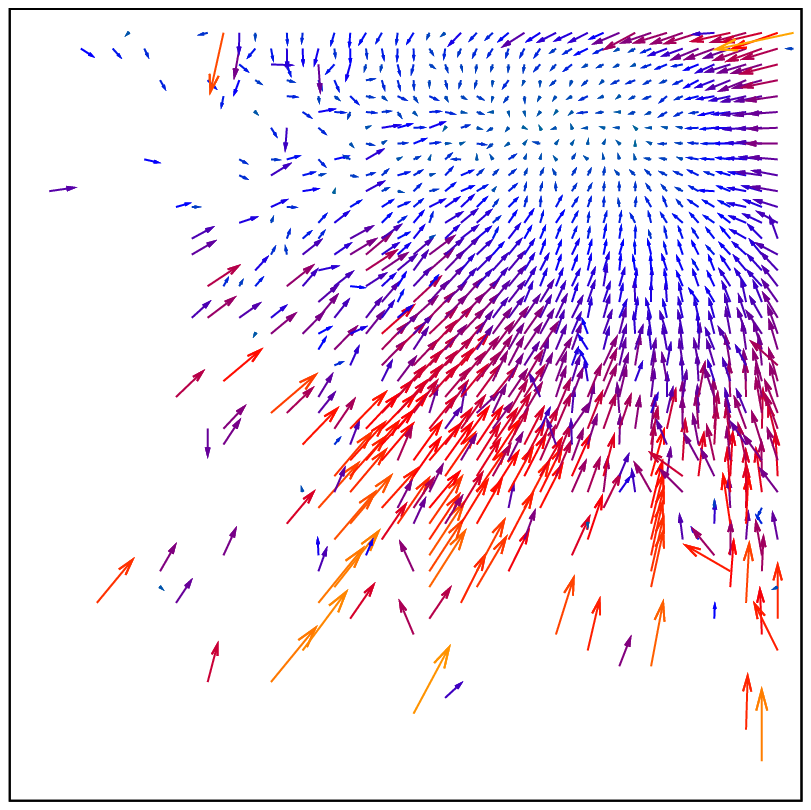}\\
\includegraphics[width=0.37\textwidth]{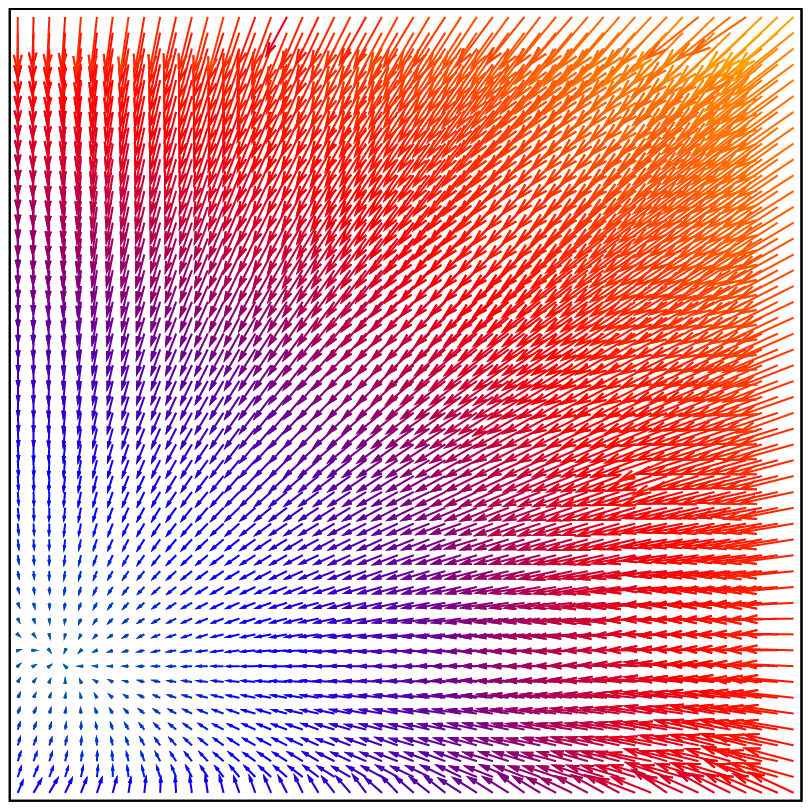}
\includegraphics[width=0.37\textwidth]{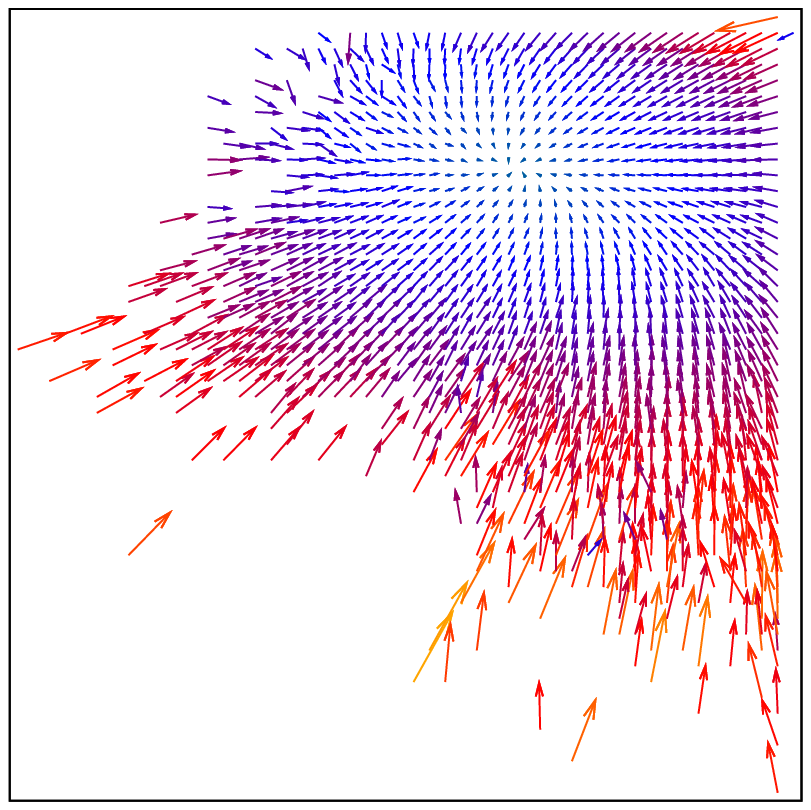}\\
\includegraphics[width=0.37\textwidth]{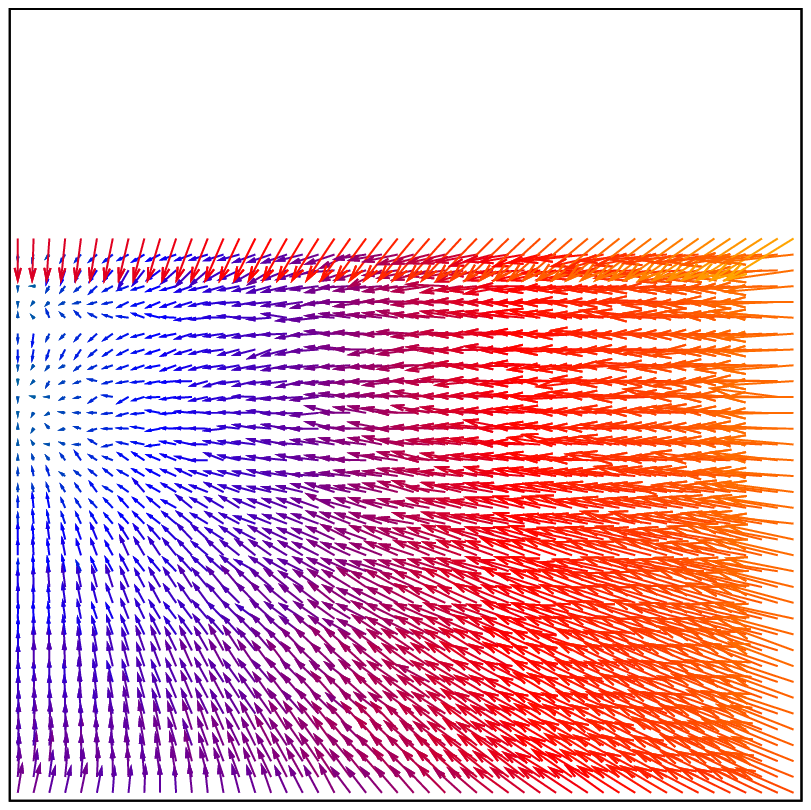}
\includegraphics[width=0.37\textwidth]{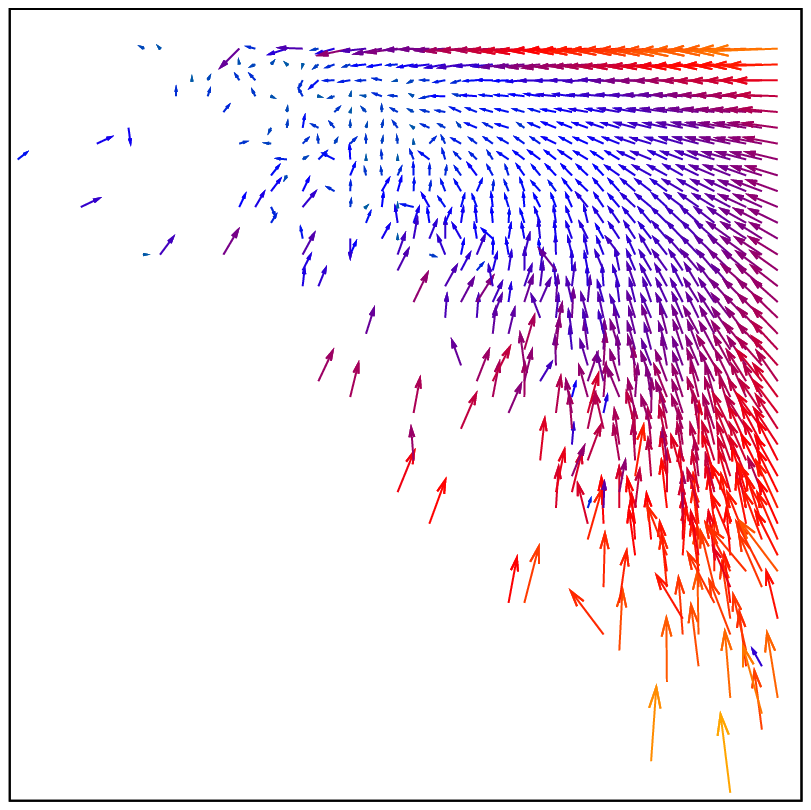}
\end{center}
\vglue 0.0cm
\caption{
Information flow on PageRank - CheiRank plane $(K,K^*)$
generated by directed links of the 
networks of Fig.~\ref{fig1}. 
Outgoing links flow is shown in linear scale $(K,K^*)$ with $K,K^*\in[1,N]$ 
on left panels, and in logarithmic scale $(\log_N K,\log_N K^*)$ 
for $\log_N K,\log_N K^*\in[0,1]$ 
on right panels.
The flow is shown by arrows which size is proportional to the
vector amplitude, which is also indicated by color 
[from yellow for large to blue for small amplitudes].
The rows corresponds to University of Cambridge (2006);
University of Oxford (2006), Wikipedia English articles,
PCN of  Linux Kernel V.2.6 (from top to bottom).
The axes show: on left column $K/N$ in   $x$-axis,
$K^*/N$ in $y$-axes; 
on right column $\log_N K$ in $x$-axis,  $\log_N K^*$ in $y$-axis;
in  all axes the variation range is $(0,1)$.
}
\label{fig9}
\end{indented}
\end{figure}

According to 2DRanking all network nodes are
distributed on a two-dimensional plane $(K,K^*)$.
The directed links of the network create an
information flow in this plane. To visualize
this flow we use the following procedure:\\
{\it a)} each node is represented by one point in the $(K,K^*)$ plane;\\
{\it b)} the whole space is divided in equal size
cells with indexes $(i,i^*)$ with the
number of nodes inside each cell being $n_{i,i^*}$,
in Fig.~\ref{fig9} we use cells of equal size in usual (left column)
and logarithmic (right column) scales;
\\
{\it c)} for each node inside the cell $(i,i^*)$, pointing to any other cell
   $(i',i^{*\prime})$, we compute the vector $(i'-i,i^{*\prime} - i^*)$ and 
average it over all nodes  $n_{i,i^*}$ inside the cell 
(the weight of links is not taken into account);\\ 
%but we checked that 
%the image of information flow is not significantly 
%modified if the weight of links according to matrix $S$ 
%is taken into account)
{\it d)} we put an arrow centered at $(i,i^*)$ with the modulus and direction
given by the average vector computed in {\it c)}.\\

Examples of such average flows for the networks of Fig.~\ref{fig1}
are shown in Fig.~\ref{fig9}. All flows have a fixed point
attractor. The fixed point is located at 
rather large values $K,K^* \sim N/4$ that is due to the fact that
in average nodes with maximal values $K,K^* \sim N$
point to lower values. At the same time
nodes with very small $K,K^* \sim 1$ still
point to some nodes which have larger values of $K,K^*$
that places the fixed point at certain intermediate
$K,K^*$ values. 
We note that the analyzed directed networks have 
dangling nodes which have no outgoing links,
the fraction of such nodes is especially large for the 
Linux network. Due to absence of outgoing links we obtain an
empty white regions in the information flow shown in Fig.~\ref{fig9}. 
A more detailed analysis of 
statistical properties of information flows on 
PageRank-CheiRank plane requires further studies.

\section{Control of spam links}

For many networks ingoing and outgoing links 
have their own  importance and thus should be 
treated on equal grounds by PageRank and CheiRank
as it is described above. However, for the WWW
it is more easy to manipulate outgoing links
which are handled by an owner of a given web page,
while ingoing links are handled by other
users. This requires to introduce some level of control on 
the outgoing links which should be taken 
into account for the ratings. Since it is very easy to create links 
to highly popular sites, we will
call ``spam links'' links for which the destination site 
is much more popular than the source.
A quantitative measure of popularity can be provided by the 
PageRank of the sites.
We do not think that spam links are frequent in  networks like 
such as procedure calls in the Linux kernel, Wikipedia and gene regulation.
Even for University networks we think that 
there are no much reasons to put spam links
inside the university domain.
However, for a large scale WWW 
an excessive number of such spam links can become harmful for 
the network performance. 
However, for WWW networks spam links are probably more widespread.
Some websites may try to improve their rating by carefully choosing 
their outgoing links. Also it is a common policy to have links 
back to a website's root pages to facilitate navigation. 
Naturally, a good rating should not be sensitive to the presence of such links.
Thus it is important to treat spam-links appropriately
in order to construct a two dimensional web-search engine.
Below we propose a method for spam links control and test it on 
an example of the Wikipedia network which has the largest size among networks
analyzed in this paper. We stress that this is done 
as a test  example and not because we think that there are 
spam links between Wikipedia articles. 
\begin{figure} 
\begin{indented}\item[]
\begin{center} 
\includegraphics[width=0.67\textwidth]{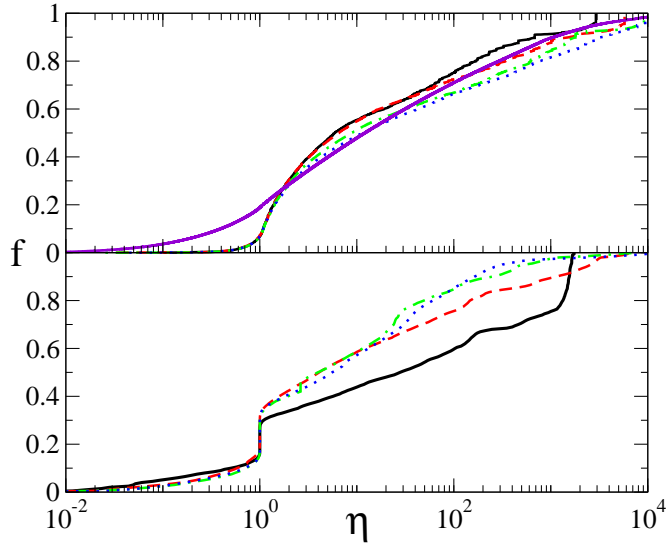}
\end{center}
\vglue 0.0cm
\caption{Fraction $f$ of inverted links as a function of 
filter parameter $\eta$
for various studied networks.
Top panel:
Wikipedia (violet curve) and
four versions of Kernel Linux PCN 
with $V2.0$ (solid black curve), $V2.3$ (dashed red curve), 
$V2.4$ (dot-dashed green curve), $V2.6$ (dotted blue curve).
Bottom panel shows data for British University networks with 
East Anglia (solid black curve), Bath (dashed red curve), 
Oxford (dot-dashed green curve), 
Cambridge 2006 (dotted blue curve).
}
\label{fig10}
\end{indented}
\end{figure} 

With this aim we propose the following filter procedure
for computation of CheiRank. The standard procedure 
described above is to invert 
the directions of all links of the network
and then to compute the CheiRank. The filter
procedure inverts a link from $j$ to $i$
only if $\eta P(K(j)) > P(K(i))$
where $\eta$ is some positive filter parameter.
After a such inversion of certain links, while other 
links remain unchanged, 
the matrix $S^*$
and $G^*$ are computed and the CheiRank vector
$P^*(K^*(i))$ of $G^*$
is determined in a usual way.
From the definition it is clear that
for $\eta=0$ there are no inverted links
and thus after filtering
$P^*$ is the same as the PageRank vector $P$.
In the opposite limit $\eta=\infty$ 
all links are inverted and $P^*$
is then the usual CheiRank discussed in previous sections. 
Thus intermediate values of $\eta$ allow to
handle the properties of CheiRank
depending on a wanted strength
of filtering. 

The dependence of the fraction  $f$ of
inverted links (defined as a ration between the number of
inverted links to the total number of links)
on the filter parameter $\eta$ is shown 
for various networks in Fig.~\ref{fig10}. 
There is a significant jump of $f$ at $\eta \approx 1$
for British University networks. In fact the condition
$\eta \approx 1$ corresponds 
approximately to the border 
relation $P(K) \approx P(K')$ with $K \approx K'$
that marks the diagonal of the $G$ matrix shown in Fig.~1
which has a significant density of matrix elements.
As a result for $\eta > 1$ we have a significant increase
of inversion of links leading to a jump of $f$
present in Fig.~\ref{fig10}. The diagonal density
is most pronounced for university networks
so that for them the jump of $f$ is mostly sharp.
 
\begin{figure} 
\begin{indented}\item[]
\begin{center}
\includegraphics[width=0.67\textwidth]{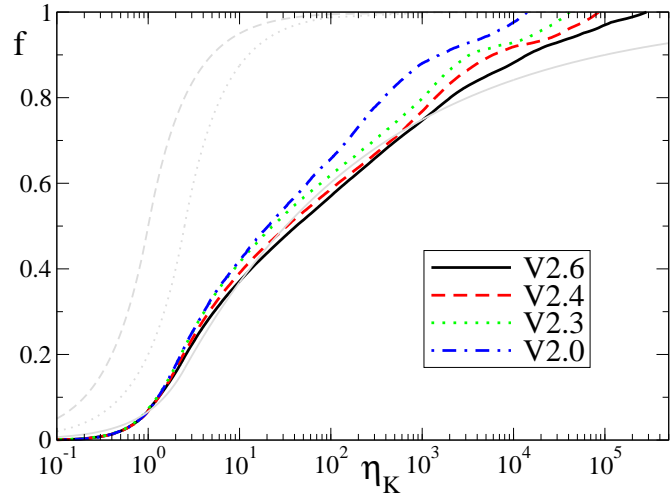}
\end{center}
\vglue 0.0cm
\caption{Fraction $f$ of inverted links in the $(K,K')$ 
plane with the condition 
$K(j) < \eta_K K(i)$ shown
as a function of 
filter parameter $\eta_K$
for Linux networks versions
shown by different curves.
Grey curves from left to right are
the theory curves with
$a=1, \nu=0$ (dashed);
$a=0.4, \nu=0$ (dotted);
$a=0.4, \nu=0.8$ (full)
(see text).
}
\label{fig11}
\end{indented}
\end{figure} 
It is also convenient to consider another condition for link inversion
defined not for $P(K_i)$ but directly in the plane $(K,K')$ 
defined by the condition: 
links are inverted only if $K(j) < \eta_K K(i)$ 
(where node $j$ points to node $i$, $j \rightarrow i$).
In a first approximation we can assume that the links 
are homogeneously distributed in the plane of transitions from $K$ to $K'$.
This density is similar to the density distribution of Google matrix elements 
$G_{K'K}$ shown in Fig.~\ref{fig1}.
For the homogeneous distribution the fraction $f$ of inverted links
is given by an area 
$\eta_K/2$ of a triangle, which height is $1$ and the basis is $\eta_K$,
for $\eta_K \leq 1$. In a similar way we have
$f=1-1/2\eta_K$ for  $\eta_K \geq 1$. We can generalize this distribution
assuming that there are only links with $1 \leq K' \leq a N$,
that is approximately the case for Linux network
where $a =0.4$ (see Fig.~\ref{fig1} bottom row),
and that inside this interval the density of links decreases 
as $1/(K')^{\nu}$. Then after computing the area we obtain the
expression for the fraction of inverted links
valid for $0 \leq \nu < 1$:
\begin{equation}   
\label{eq3}                                                                          
f(\eta_K)=\left\{\begin{array}{cr}                                       
\frac{1-\nu}{2-\nu} (a \eta_K) & \eta_K\leq1/a\\             
1+\left(\frac{1-\nu}{2-\nu}-1\right)\left(a\eta_K\right)^{\nu-1} \ \ \ \ \   
& \eta_K>1/a                    
\end{array}                                                                                
\right.                                 
\end{equation}
The comparison of this theoretical expression 
with the numerical data for Linux PCN is shown in Fig.~\ref{fig11}.
It shows that the data for Linux are well described by the theory
(\ref{eq3}) with $a=0.4$ and $\nu=0.8$. The last value 
takes into account the fact that
the density of links decreases with PageRank index $K'$
as it is well visible in Fig.~\ref{fig1}.

\begin{figure} 
\begin{indented}\item[]
\begin{center}
\includegraphics[width=0.37\textwidth]{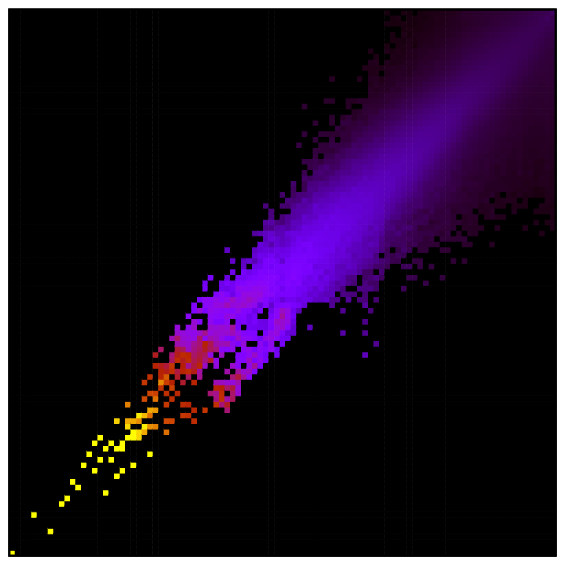}
\includegraphics[width=0.37\textwidth]{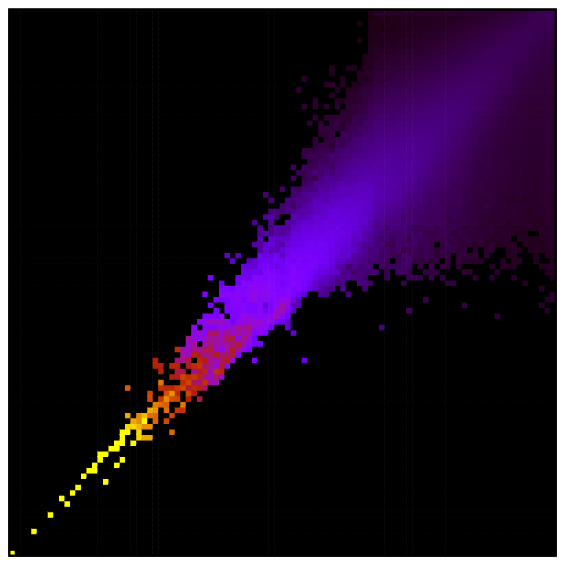}\\
\includegraphics[width=0.37\textwidth]{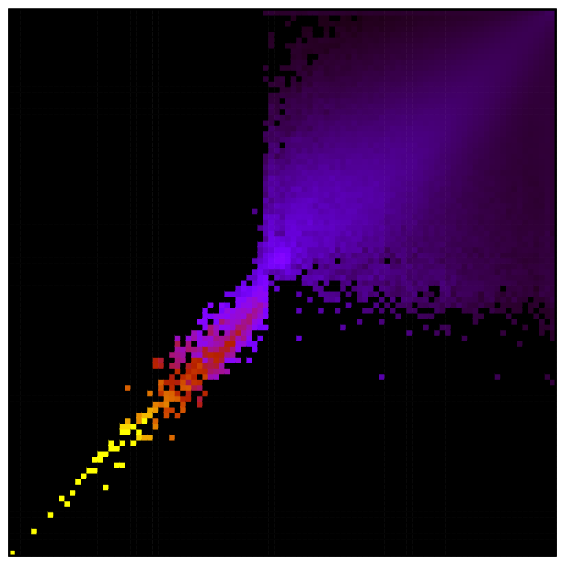}
\includegraphics[width=0.37\textwidth]{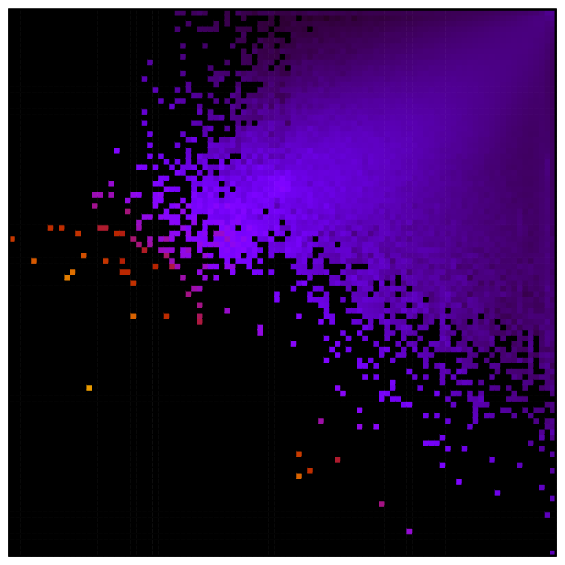}
\end{center}
\vglue 0.0cm
\caption{Density distribution $W(K,K^{*})=dN_i/dKdK^*$ for Wikipedia 
in the plane of PageRank and 
filtered CheiRank indexes, $(\log_N K,\log_N K^{*})$, 
in a equidistant $100\times100$ lattice with $\log_N K,\log_N K^*\in[0,1]$. 
The filter parameter is $\eta=10$ (left-top panel),
$100$ (right-top panel), $1000$ (left-bottom panel),
$10^5$ where 
all links are inverted (right-bottom panel).
The color panel is the same as in Fig.~\ref{fig3} with 
the saturation value
$W_s^{1/4}=0.5W_M^{1/4}$.
The axes show: $\log_N K$ in $x$-axis,  $\log_N K^*$ in $y$-axis,
in both axes the variation range is $(0,1)$.
}
\label{fig12}
\end{indented}
\end{figure} 

The variation of nodes density in the plane of
PageRank and filtered CheiRank $(K,K^*)$
for the Wikipedia network is shown in Fig.~\ref{fig12}
with the filtering by $\eta$ for $P(K)$ and $P(K')$ values.
At moderate values $\eta=10$ the density
is concentrated near diagonal,
with further increase of $\eta=100; 1000$
a broader density distribution appears
at large $K$ values which goes to smaller and smaller
$K$ until the limiting distribution 
without filtering is established at very large $\eta$.
The top 100 Wikipedia articles obtained with filtered CheiRank
at the above values of $\eta$ are given at
\cite{dvvadi}. We also give there top articles in
2DRank which gives articles in order of their
appearance on the borders of a square of increasing size
in $(K,K^*)$ plane (see detailed description in \cite{wiki}).
These data clearly show that filtering eliminates
articles with many outgoing links and gives 
a significant modification
of top CheiRank articles. Thus the described method can be 
efficiently used for control of spam links
present at the WWW.

\section{2DRanking of gene regulation networks}

The method of 2DRanking described above 
is rather generic and can be applied to various types
of directed networks. Here we apply it to 
gene regulation networks of Escherichia Coli and Yeast
with the network links taken from \cite{ualon}.
Such transcription regulation networks control
the expression of genes and have important biological 
functions \cite{gene}.

\begin{figure}
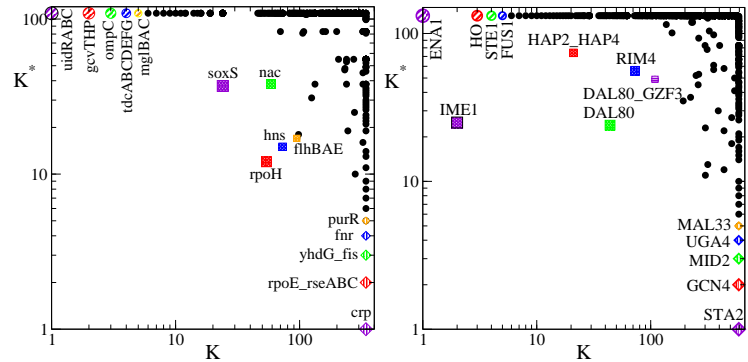
 
\begin{indented}\item[]
\begin{center}
  \includegraphics[width=0.37\textwidth]{fig13a.eps}
 \includegraphics[width=0.37\textwidth]{fig13b.eps}
\end{center}
\vglue 0.0cm
\caption{Distribution of nodes in the plane of PageRank $K$ 
and CheiRank $K^*$ for Escherichia Coli, and Yeast transcription networks on
left and right panels respectively
(network data are taken from \cite{ualon}). 
The nodes with five top probability values of PageRank, 
CheiRank and 2DRank 
are labeled by their corresponding node names; 
they correspond to 5 lowest index values.
}
\label{fig13}
\end{indented}
\end{figure} 

The distribution of nodes in PageRank-CheiRank plane is shown 
in Fig.~\ref{fig13}. The  top 5 nodes in CheiRank probability value
(lowest CheiRank indexes)
are those which send many outgoing orders,
top 5 in PageRank probability are those which obtain
many incoming signals and the top 5 indexes in 2DRank
(with 5 lowest 2DRank index values)
combine these two functions. For these networks the correlator $\kappa$
is close to zero (even slightly negative)
which indicates the statistical independence between outgoing and ingoing
links quite similarly to the case of the PCN for the Linux Kernel.
This may indicate that a slightly negative correlator $\kappa$ 
is a generic property for the data flow network of control and regulation 
systems. Whether the obtained ratings can bring some insights 
on the functioning of gene regulation 
can only be assessed by experts in the field. However, we hope that 
such an analysis will prove to be useful for a better understanding 
of gene regulation networks. 

\section{Discussion}

Above we presented extensive studies of statistical properties
of 2DRanking based of PageRank and CheiRank
for various types of directed networks.
All studied networks are of a free-scale type with
an algebraic distribution of ingoing and outgoing
links with a usual values of exponents.
In spite of that their statistical characteristics
related to  PageRank and CheiRank are rather different.
Some networks have high correlators between PageRank
and CheiRank (e.g. Wikipedia, British Universities),
while others have practically zero correlators
(PCN of Linux Kernel, gene regulation networks).
The distribution of nodes in PageRank-CheiRank plane
also varies significantly between different types of networks.
Thus 2DRanking discussed here gives more 
detailed  classification
of information flows on directed networks.
\begin{figure} 
\begin{indented}\item[]
\begin{center}
\includegraphics[width=0.37\textwidth]{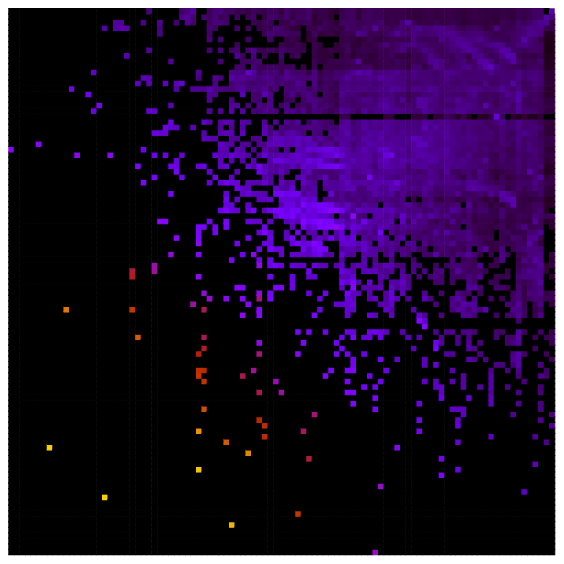}
\includegraphics[width=0.37\textwidth]{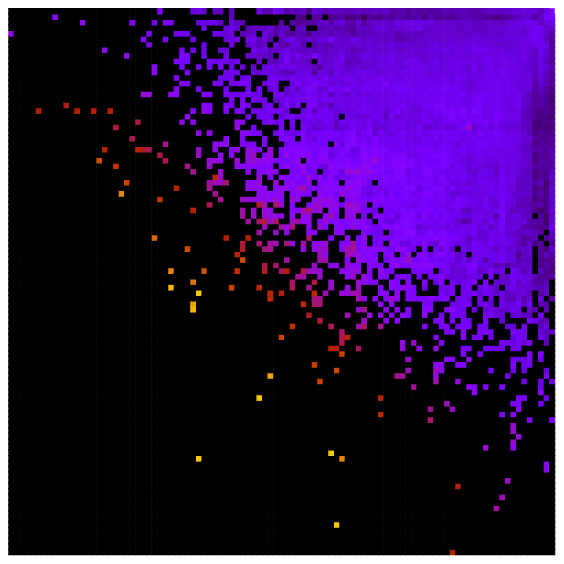}\\
\includegraphics[width=0.37\textwidth]{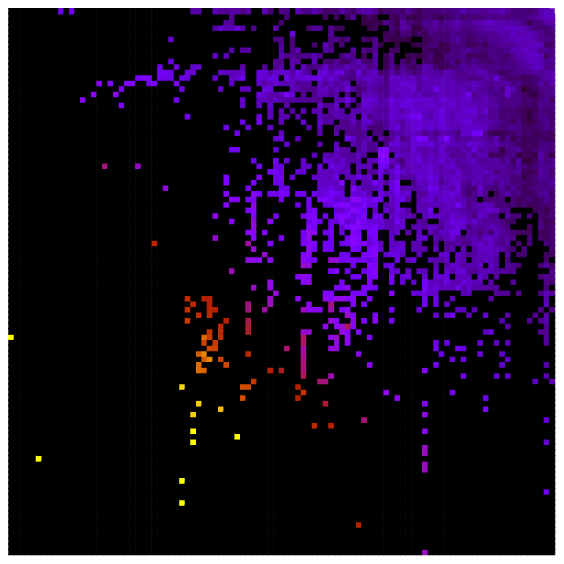}
\includegraphics[width=0.37\textwidth]{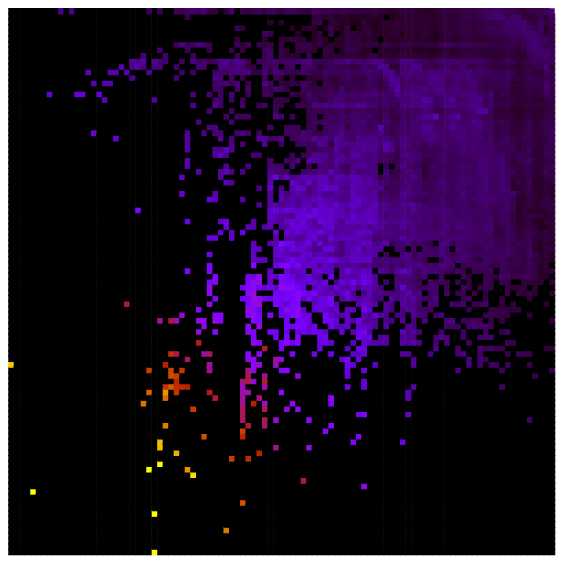}
\end{center}
\vglue 0.0cm
\caption{Density distribution $W(K,K^{*})=dN_i/dKdK^*$ 
shown in the same frame as in Fig.~\ref{fig3}
for networks collected in 2011:
University of Cambridge (top left),
University of Bologna (top right),
ENS Paris for crawling level 5 (bottom left)
and 7 (bottom right).
The color panel is the same as in Fig.~\ref{fig3} with 
the saturation value
$W_s^{1/4}=0.5W_M^{1/4}$.
The axes show: $\log_N K$ in $x$-axis,  $\log_N K^*$ in $y$-axis,
is both axes the variation range is $(0,1)$.
}
\label{fig14}
\end{indented}
\end{figure} 

\begin{figure} 
\begin{indented}\item[]
\begin{center}
\includegraphics[width=0.67\textwidth]{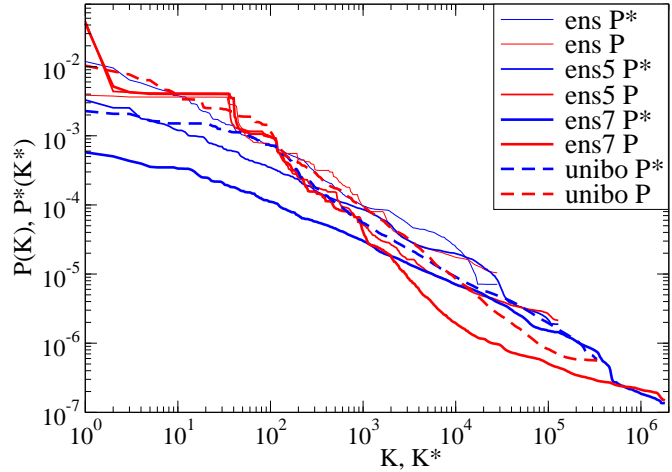}
\end{center}
\vglue -0.0cm
\caption{Dependence of probabilities of
PageRank $P(K)$ (red curve) and CheiRank $P^*(K^*)$ (blue curve)
on corresponding ranks $K$ and $K^*$
for the networks of ENS Paris 
(crawling levels 3,5,7) and University of Bologna.
}
\label{fig15}
\end{indented}
\end{figure}

We think that 2DRanking gives new possibilities for 
information retrieval from large databases which 
are growing rapidly with time. Indeed, for example 
the size of the Cambridge network increased
by a factor 4 from 2006 to 2011 (see 
Appendix and Fig.~\ref{fig2}). At present, 
web robots start automatically generate 
new webpages. These features can be responsible for
appearance of gaps in density distribution in
$(K,K^*)$ plane at large $K, K^* \sim N$ values
visible for large scale university
networks of Cambridge and ENS Paris in 2011 
(see Fig.~\ref{fig14}). Such an automatic generation of links
can change the scale-free properties of networks.
Indeed, for  ENS Paris we observe appearance of 
large step in the PageRank distribution 
$P(K)$ shown in Fig.~\ref{fig15}.
This step for $P(K)$ remains not sensitive to the deepness
of crawling which goes on a level of $3, 5$ and $7$ links.
However, the CheiRank distribution
changes with the deepness level
becoming more and more flat (see  Fig.~\ref{fig15}).
Such a tendency in a modification of 
network statistical properties 
is visible in 2011 for large
size university networks, while
networks of moderate size, 
like University of Bologna 2011
(see data in Figs.~\ref{fig14},\ref{fig15}), 
are not yet affected. A sign of ongoing changes is 
a significant growth of the correlator
value $\kappa$ which increases
up to very large value ($30$ for Cambridge 2011 
and $63$ for ENS Paris).
There is a danger that automatic generation
of links can lead to a delocalization transition
of PageRank that can destroy
efficiency of information retrieval from the WWW.
We note that it is known that PageRank delocalization
can appear in certain models of Markov chains
and Ulam networks \cite{ulamnetwork}.
Such a delocalization of PageRank would 
make the ranking of nodes inefficient 
due to high sensitivity of ranking to fluctuations
that would create a very dangerous situation for
the WWW information retrieval and ranking.
We also note that the spectrum of the Google matrix
of British universities networks
has been recently analyzed in \cite{univpagerank}.
The spectrum and eigenstates analysis can be a sensitive
tool for location of precursors of a delocalization transition. 

\begin{figure} 
\begin{indented}\item[]
\begin{center}
\includegraphics[width=0.8\textwidth]{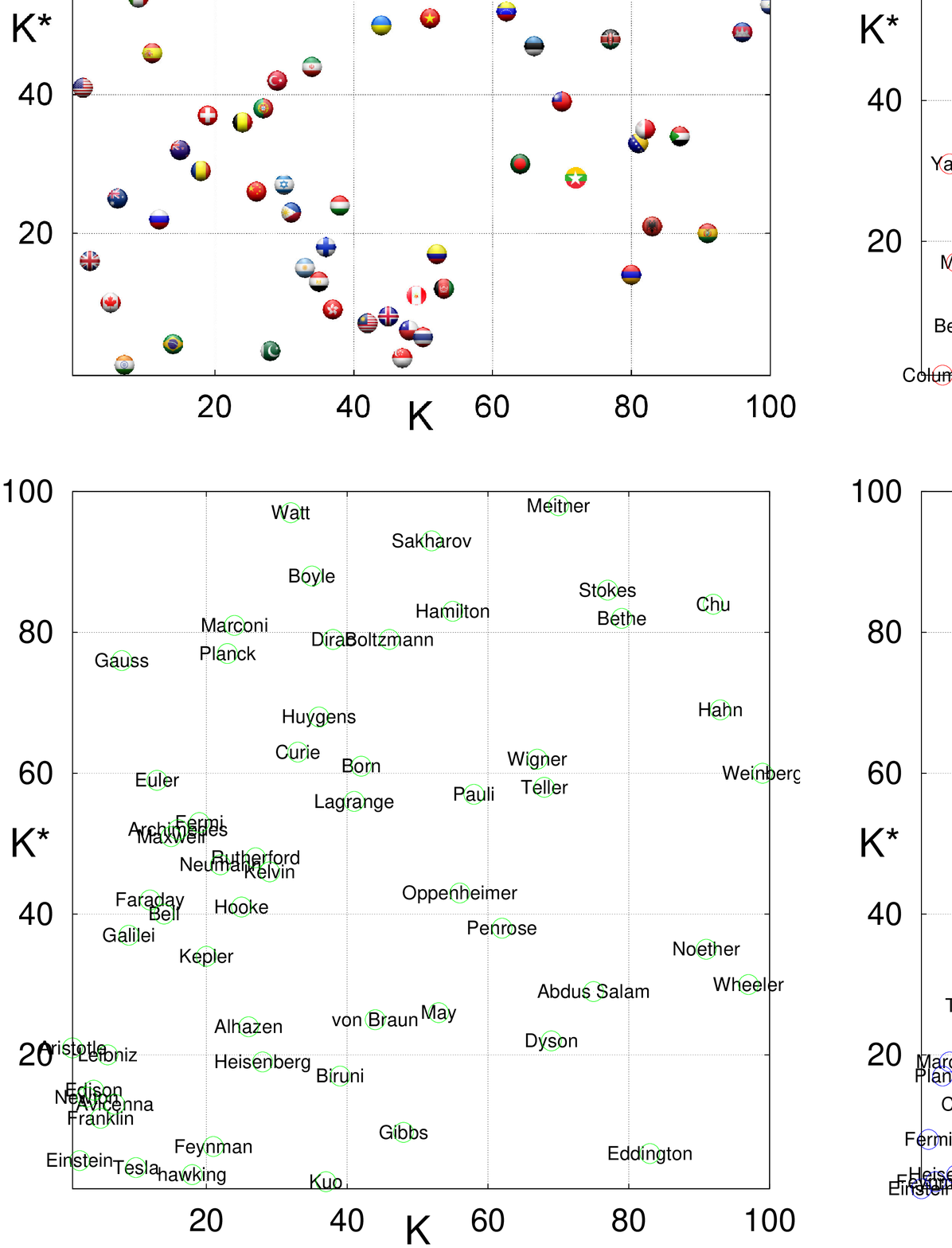}
\end{center}
\vglue -0.0cm
\caption{Examples of Dvvadi search analysis
of Wikipedia articles shown on 2D plane of
PageRank  $K$ and CheiRank $K^*$ 
local indexes for specific subjects (articles): 
countries marked by their flag (top left), universities (top right),
physicists (bottom left),
Nobel laureates in physics (bottom right), circles mark the node location; 
high resolution figures and listings of
names with local $(K,K^*)$ values in $100 \times 100$
square are available at \cite{dvvadi}
(listings with global ranking are available at \cite{wiki}).
}
\label{fig16}
\end{indented}
\end{figure}

Our studies of 2DRanking pave the way to development of
two-dimensional search engines which will use the 
advantages of both PageRank and CheiRank.
Indeed, the Google search engine uses 
as the fundamental mathematical basis 
the one-dimension ranking
related to PageRank \cite{meyerbook}.
Of course, there are various other important elements
used by the Google search which  remain
the company secret and not only PageRank
order matters for the Google ranking.
However, the mathematical aspects of these additional
elements are not really known
(e.g. they are not described in  \cite{meyerbook}).
At the same time the size of databases
generated by the modern society
continues its enormous growth.
Due to that the information retrieval
and ordering of such data sets 
becomes of primary importance
and new mathematical tools should be developed
to operate and characterize 
efficiently their information flows and ranking.
Here we proposed and analyzed the properties
 the new two-dimensional
search engine, which we call {\bf Dvvadi}
from Russian ``dva (two)'' and ``dimension'',
will use the complementary ranking abilities
of both PageRank and CheiRank.
Now the procedure of ordering of all
network nodes uses not one but two vectors
of the Google matrix of a network.
The computational efforts are twice more expensive
but for that we obtain a new quality 
since now the nodes are ranked 
in 2D plane not only by
their degree of popularity but also by their
degree of communicability.
Thus for the Wikipedia network the
top 3 articles in PageRank probability are
3 countries (most popular), 
while the top 3 articles in CheiRank probability
are 3 listings of knowledge, state leaders and geographical places
(most communicative).
Hence, we can rank the nodes of the network in 
a new two-dimensional manner  which highlight
complementary properties of node popularity and communicability.
Thus the Dvvadi search can present nodes not in a line but on a 2D plane
characterizing these two complementary properties of nodes.
Examples of such 2D representation of nodes selected from Wikipedia
articles
by a specific subject are shown in Fig.~\ref{fig16}:
we determine global $K$ and $K^*$ indexes of all articles,
select a specific subject (e.g. {\it countries})
and then represent countries in the local index $K$ and $K^*$
corresponding to their appearance in the global order
via PageRank and CheiRank. For countries we see a 
clear tendency that the countries on the top
of PageRank probability (low $K$) have relatively high
CheiRank index (high $K^*$) (e.g. US, UK, France)
while small countries in 
the region $K \approx 50, K^* \approx 10$
have another tendency (e.g. Singapore).
We attribute this to specific
routes of cultural and industrial development of the world:
e.g. Singapore was a colony of UK became
a strong trade country and due to that
have historically many links pointing to
UK and other developed countries.
For universities we also see that
those at the top of PageRank (Harvard, Oxford, Cambridge)
are not very communicative having high $K^*$ values,
while Columbia, Berkeley are more balanced and Florida
 and FSU are very communicative
probably due to initial location of Wikimedia Foundation at Florida.
For physicists
we see that links to many scientific fields
(like Shen Kuo) or polularization of science
(like Hawking and Feynman) place those
people at the top positions of CheiRank.
In a similar way for the
Nobel laureates in physics 
we see that CheiRank stresses the communicative
aspects: e.g. Feynman, due to his
popularization of physics;
Salam due to the Institute on his name at 
Trieste with a broad international activity;
Raman due to Raman effect. 

On the basis of the above results
we think that PageRank-CheiRank classification of network
nodes on 2D plane will allow to analyze the information flows
on directed networks in a better way. 
It is also important to note that 2DRanking is very
natural for financial and trade networks.
Indeed, the world trade usually uses the import and export ranking which
is analogous to PageRank and CheiRank, as it is shown in \cite{wtrade}.
We think that such {\bf Dvvadi} engine/motor \cite{dvvadi}
will find useful applications for 
treatment of enormously large
databases created by modern society.

% {\bf Acknowledgments:} 
\ack
We thank K.M.Frahm $\;$  and  $\;\;$ B.Georgeot
for useful discussions of properties of British University networks.
This work is done in the frame of the EC FET Open project 
``New tools and algorithms for directed network analysis''
(NADINE $No$ 288956).

\renewcommand{\theequation}{A-\arabic{equation}}
  % redefine the command that creates the equation no.
  \setcounter{equation}{0}  % reset counter
\renewcommand{\thefigure}{A-\arabic{figure}}
  % redefine the command that creates the figure no.
  \setcounter{figure}{0}  % reset counter

\appendix
\section{Appendix}

We list below the directed networks used in this work
giving for them 
number of nodes $N$, number of links $N_{links}$
and correlator between PageRank and CheiRank $\kappa$.
Additional data can be find at \cite{dvvadi}.

{\bf Linux Kernel  Procedure Call Networks} are taken from \cite{alik}
(see also \cite{wlinux}) with the parameters for various
kernel versions  shown in Table \ref{table1}\\

\begin{table}[h]%
\caption{\label{table1} Linux Kernel  network parameters}
\begin{indented}
\item[]
\begin{tabular}{cccc} 
\br
version& $N$&$N_{links}$&$\kappa$\\
\mr
{\bf V1.0}&2752& 5933& $\kappa=-0.11$\\
{\bf V1.1}&4247& 9710& $\kappa=-0.083$\\
{\bf V1.2}& 4359& 10215& $\kappa=-0.048$\\
{\bf V1.3}& 10233& 24343& $\kappa=-0.102$\\
{\bf V2.0}& 14080& 34551& $\kappa=-0.037$\\
{\bf V2.1}& 26268& 59230& $\kappa=-0.058$\\
{\bf V2.2}& 38767& 87480& $\kappa=-0,022$\\
{\bf V2.3}& 41117& 89355& $\kappa=-0.081$\\
{\bf V2.4}& 85757& 195106& $\kappa=-0.034$\\
{\bf V2.6}& 285510& 588861& $\kappa=0.022$\\
\br
\end{tabular}
\end{indented}
\end{table}

Web networks of {\bf British Universities} dated by year 2006 are 
taken from \cite{britishuniv}, and are shown in Table \ref{table2}.

\begin{table}[h]%
\caption{\label{table2} British Universities network parameters}
\begin{indented}
\item[]
\begin{tabular}{cccc} 
\br
University& $N$&$N_{links}$&$\kappa$\\
\mr
{\bf RGU (Abardeen)}&  1658& 15295 & $\kappa=1.03$\\
{\bf Uwic (Wales)}& 5524& 111733& $\kappa=0.82$\\
{\bf NTU (Nottingham)}& 6999& 143358& $\kappa=0.50$\\
{\bf Liverpool}&  11590& 141447& $\kappa=1.49$\\
{\bf Hull}& 16176& 236525& $\kappa=5.31$\\
{\bf Keele}& 16530& 117944& $\kappa=3.24$\\
{\bf UCE (Birmingham)}& 18055 &351227& $\kappa=1.67$\\
{\bf Kent}& 31972 &277044& $\kappa=2.65$\\
{\bf East Anglia}& 33623& 325967& $\kappa=5.50$\\
{\bf Sussex}& 54759& 804246& $\kappa=7.29$\\
{\bf York}& 59689& 414200& $\kappa=8.13$\\
{\bf Bath}& 73491& 541351& $\kappa=3.97$\\
{\bf Glasgow}& 90218& 544774& $\kappa=2.22$\\
{\bf Manchester}& 99930& 1254939& $\kappa=3.47$\\
{\bf UCL (London)}& 128450& 1397261& $\kappa=2.33$\\
{\bf Oxford}& 200823& 1831542& $\kappa=4.66$\\
{\bf Cambridge (2006)}& 212710& 2015265& $\kappa=1.71$\\
\br
\end{tabular}
\end{indented}
\end{table}

We also developed a special code with which we performed crawling of
university web networks in January - March 2011 with the parameters given below:
{\bf University of Cambridge (2011)} 
with $N=898262$, $N_{links}=15027630$, $\kappa=30.0$;
{\bf \'Ecole Normale Sup\'erieure, Paris (ENS 2011)} with 
$N=28144$, $N_{links}=971856$, $\kappa=1.67$ 
(crawling deepness level of 3 links),
$N=129910$, $N_{links}=2111944$, $\kappa=16.2$ 
(crawling deepness level of 5 links),
$N=1820015$, $N_{links}=25706373$, $\kappa=63.6$ 
(crawling deepness level of 7 links); 
{\bf University of Bologna}
with $N=339872$, $N_{links}=16345488$, $\kappa=2.63$.

The data for hyperlink network of {\bf Wikipedia English articles (2009)}
are taken from \cite{wiki}
with $N=3282257$, $N_{links}=71012307$, $\kappa=4.08$.

{\bf Transcription Gene} networks are taken from \cite{ualon}. We have for them:
{\bf Escherichia Coli} with $N=423$, $N_{links}=519$, $\kappa=-0.0645$;
{\bf Yeast} with $N=690$, $N_{links}=1079$,  $\kappa=-0.0497$;
for all links the weight is take to be the same.

{\bf Business Process Management} network is taken from
\cite{business} with $N=175$, $N_{links}=240$, $\kappa=0.164$.

{\bf Brain Model} network is taken from \cite{brain} with
$N=10000$, $N_{links}=1960108$, 
$\kappa=-0.054$ (unweighted), $\kappa=-0.065$ (weighted).

\end{document}